\documentclass[preprint,  12pt]{aastex62}
\usepackage[T1]{fontenc}

\usepackage{bm}        
\usepackage{amssymb, amsmath}   
\usepackage{cancel}
\usepackage{color}

\newcommand{\vect}[1]{\bm{#1}}

\renewcommand{\thefootnote}{\arabic{footnote}}

\def\apj{Astrophysical Journal}                 
\def\apjl{Astrophysical Journal}                 
\def\mnras{MNRAS}                 
\def\ssr{Space Science Reviews}                 
\def\aap{Astronomy and Astrophysics}                 
\DeclareMathOperator{\sech}{sech}

\begin{document}



\title{Magnetic Energy Release, Plasma Dynamics and Particle Acceleration in Relativistic Turbulent Magnetic Reconnection}

\author{Fan Guo}
\affiliation{Los Alamos National Laboratory, NM 87545 USA}
\author{Xiaocan Li}
\affiliation{Dartmouth College, Hanover, NH 03750 USA}
\author{William Daughton}
\affiliation{Los Alamos National Laboratory, NM 87545 USA}
\author{Hui Li}
\affiliation{Los Alamos National Laboratory, NM 87545 USA}
\author{Patrick Kilian}
\affiliation{Los Alamos National Laboratory, NM 87545 USA}
\author{Yi-Hsin Liu}
\affiliation{Dartmouth College, Hanover, NH 03750 USA}
\author{Qile Zhang}
\affiliation{Los Alamos National Laboratory, NM 87545 USA}
\author{Haocheng Zhang}
\affiliation{Purdue University, West Lafayette, IN 47907, USA}
\date{\today}

\begin{abstract}
In strongly magnetized astrophysical plasma systems, magnetic reconnection is believed to be the primary process during which explosive energy release and particle acceleration occur, leading to significant high-energy emission. Past years have witnessed active development of kinetic modeling of relativistic magnetic reconnection, supporting this magnetically dominated scenario. A much less explored issue in studies of relativistic reconnection is the consequence of three-dimensional dynamics, where turbulent structures are naturally generated as various types of instabilities develop. This paper presents a series of three-dimensional, fully-kinetic simulations of relativistic turbulent magnetic reconnection (RTMR) in positron-electron plasmas with system domains much larger than kinetic scales. Our simulations start from a force-free current sheet with several different modes of long wavelength magnetic field perturbations, which drive additional turbulence in the reconnection region. Because of this, the current layer breaks up and the reconnection region quickly evolves into a turbulent layer filled with coherent structures such as flux ropes and current sheets.  We find that plasma dynamics in RTMR is vastly different from their 2D counterparts in many aspects. The flux ropes evolve rapidly after their generation, and can be completely disrupted due to the secondary kink instability. This turbulent evolution leads to superdiffusion behavior of magnetic field lines as seen in MHD studies of turbulent reconnection.  Meanwhile, nonthermal particle acceleration and energy-release time scale can be very fast and do not strongly depend on the turbulence amplitude. The main acceleration mechanism is a Fermi-like acceleration process supported by the motional electric field, whereas the non-ideal electric field acceleration plays a subdominant role. We also discuss possible observational implications of three-dimensional RTMR in high-energy astrophysics.
\end{abstract}

\keywords{}


\section{Introduction} Magnetic reconnection is one of the fundamental plasma processes in the universe where free magnetic energy stored in the anti-parallel magnetic field component of current sheets can rapidly release and be converted into energies contained in plasma bulk flow, and thermal and nonthermal distributions \citep{Biskamp2000,Priest_2007}. In strongly magnetized astrophysical systems, magnetic reconnection is an efficient mechanism for converting magnetic energy into particle heating and acceleration, and subsequent high-energy emissions.
Relativistic magnetic reconnection in the magnetically dominated regime (magnetization parameter $\sigma=B^2/(8\pi w)\gg1$, where $w$ is the enthalpy density) is often invoked to explain high-energy particles and emissions from objects such as pulsar wind nebulae \citep{Coroniti1990,Lyubarsky2001,Kirk2003,Arons2012,Hoshino2012}, jets from active galactic nuclei \citep{Pino2005,Giannios2009,Zhang2015,Yan2015,Zhang2017,Zhang2018,Nathanail2020,Zhang2021a} and gamma-ray bursts \citep{Zhang2011,McKinney2012}. 

Past years have seen an active development on theoretical modeling of relativistic magnetic reconnection, supporting the magnetically dominated scenario \citep{Blackman1994,Lyutikov2003,Lyubarsky2005,Comisso2014,Liu2017,Liu2020}. Recently, a large range of two-dimensional (2D), fully-kinetic studies
have intensely explored reconnection in the magnetically dominated regime on the reconnection properties \citep{Liu2015,Liu2017,Liu2020}, particle acceleration \citep{Sironi2014,Guo2014,Guo2015,Guo2016,Guo2019,Werner2016}, and possible radiation signatures \citep{Sironi2016,Petropoulou2016,Zhang2018,Zhang2020}. In the magnetically dominated regime, reconnection in a thin current sheet proceeds at a rate with inflow speed $v_{in} \sim0.1$ times of the upstream Alfv\'en speed $V_A$ close to the maximum local rate \citep{Liu2017,Liu2020}. While the maximum outflow Lorentz factor can approach $\Gamma_{out} \sim \sqrt{\sigma+1}$, it can be significantly limited by a guide field $B_g$ so $\Gamma_{out} \sim \sqrt{(\sigma+1)/(\sigma_g+1)}$, where $\sigma_g =B_g^2/(8\pi w)$ \citep{Liu2015}.
Relativistic reconnection seems to be a source of efficient nonthermal particle acceleration \citep{Sironi2014,Guo2014,Guo2015,Werner2016}. In 2D simulations, it was found, through several different analysis, that a Fermi-like acceleration driven by plasmoid motion dominates the acceleration process \citep{Guo2014,Guo2015,Guo2019}\footnote{see \citep{Drake2006,Dahlin2014,Dahlin2017,Li2017,Li2018Role,Li2019Particle,Li2019Formation} for a nonrelativistic description} and leads to formation of power-law energy distributions $f \propto \varepsilon^{-p}$ in the weak guide field regime. In particle-in-cell (PIC) simulations of weak-guide-field relativistic reconnection, the plasma dynamics and particle acceleration in 2D relativistic reconnection is controlled by plasmoids as current sheets continuously generate and break up into interacting plasmoids \citep{Daughton2007,Guo2015,Sironi2016,Liu2020}. See \citet{Guo2020} for a review on the primary acceleration mechanism, power-law formation and reconnection physics.

While 2D anti-parallel relativistic reconnection simulations have been extensively carried out, there have been only limited studies on 3D magnetically dominated reconnection with $\sigma \gg 1$ \citep{Sironi2014,Guo2014,Guo2015,Werner2017}. It was verified that 3D physics does not strongly influence the development of nonthermal power-law energy spectrum\footnote{however, see \citet{Li2019Formation,Zhang2021b} for the effect of 3D physics on nonthermal particle acceleration in nonrelativistic reconnection}, but the acceleration mechanism has not been studied as carefully as in 2D simulations. In addition, the time scale of energy release represented by reconnection rate \citep{Guo2015} does not strongly change in 3D\footnote{see also nonrelativistic studies \citep{Liu2013,Daughton2014}}. However, there has been a lack of exploration on 3D plasma dynamics and its possible observational implications.  It is known now that a range of secondary instabilities can grow and lead to turbulence in the reconnection layer  \citep{Bowers2007,Zenitani2008,Daughton2011,Guo2015,Kowal2020}. Therefore, turbulent magnetic reconnection is a more likely picture for the realistic situation and conclusions based on 2D reconnection need to be carefully examined using 3D simulations. 

To summarize what we have discussed so far: a major uncertainty of the relativistic reconnection studies is the role of three-dimensional dynamics and turbulence. Earlier turbulent reconnection studies using various numerical approaches have found that pre-existing turbulence and self-generated turbulence can both exist, but their roles on magnetic energy dissipation, plasma dynamics, and particle acceleration remain controversial \citep{Bowers2007,Huang2016,Beresnyak2017,Kowal2017,Lazarian1999,Matthaeus1985,Matthaeus1986,Loureiro2009,Daughton2011,Daughton2014,Yang2020,Leake2020}. This highlighted study is designed to explore some of the aspects when magnetic reconnection occurs in a turbulent state, with focus on the relativistic regime, which we now refer to as relativistic turbulent magnetic reconnection (RTMR).  

In this paper, a number of large-scale (system size $\gg$ kinetic scale), three-dimensional, fully-kinetic simulations are carried out using the Los Alamos VPIC code\footnote{https://github.com/lanl/vpic}. This is made possible on the Trinity machine during its open science period. We focus on simulations with positron-electron pair plasma that minimize the kinetic range and maximize the ratio between system size and kinetic scale ($L/d_e \sim 10^3$). Different from typical kinetic studies, we have added a new set of initial  perturbations to drive extra turbulence in the simulation domain. Because of this, the reconnection layer quickly develops into a turbulent state. The flux ropes, different from their corresponding 2D magnetic islands, evolve dynamically after their generation, and can be completely disrupted due to the secondary kink instability. We find that while the reconnection X-points are strongly perturbed by the injected and self-generated fluctuation due to secondary tearing and kink instabilities, acceleration of high-energy particles is robust and leads to formation of power-law distribution. In addition, the normalized reconnection rate is on the order of $R \sim 0.1$. We show that for the anti-parallel reconnection case the Fermi-like acceleration mechanism is the dominant process. During the reconnection process, thin reconnection layers continuously develop and the peak reconnection rate is nearly independent of the injected magnetic energy, suggesting the reconnection physics is primarily controlled by kinetic-scale physics in kinetic our simulations. The rest of the paper is organized as follows: Section 2 describes numerical methods, setups and important parameters. Section 3 discusses the primary simulation results. We discuss observational implications in Section 4, and conclusions are made in Section 5.

\section{Numerical simulations} The 3D simulations presented here start from a force-free current layer with $\vect{B}=B_0\tanh(z/\lambda)\hat{x} +
B_0\sqrt{\sech^2(z/\lambda) + b_g^2}\hat{y}$, where $B_0$ is the strength of the
reconnecting magnetic field, $b_g$ is the strength of the guide field $B_g$ normalized by
$B_0$, and $\lambda$ is the half-thickness of the current sheet \citep{Guo2014,Guo2015,Guo2019,Li2017,Li2018Role,Li2019Formation}. 
The plasma consists of 
electron-positron pairs (mass ratio $m_p/m_e=1$). The initial distributions are Maxwellian with a uniform density $n_0$ 
and temperature ($T_{p}=T_{e}$). For the simulations discussed in this paper, the amount of thermal energy per particle is $\sim m_ec^2$.
Particles in the sheet have a drift velocity $\textbf{u}_p=-\textbf{u}_e$, and that gives rise to a current density satisfying Ampere's law $\nabla\times\textbf{B}=4\pi\textbf{J}$. The simulations are
performed using the VPIC code \citep{Bowers2008}, which solves the relativistic Vlasov-Maxwell equation system. We have performed simulations with $\sigma_e = B^2_0/(4 \pi n_e m_e c^2) = 6 - 1600$. We mainly focus on the case with $\sigma_e=100$, corresponding to $\omega_{pe}/\Omega_{ce}=0.1$, where $\omega_{pe} = \sqrt{4 \pi n_e e^2/m_e}$ is the plasma frequency and $\Omega_{ce} = eB_0/(m_e c)$ is the electron gyrofrequency (without relativistic corrections). In the simulations the half-thickness is set to be $\lambda=6d_e$ for simulations with $\sigma_e < 100$, $\lambda=12d_e$ for $\sigma_e = 400$ and $\lambda=24d_e$ for $\sigma_e = 1600$ to ensure drift speed $u_i < c$. The presented electric and magnetic fields are normalized by $B_0$. The current density is normalized by $J_0 = e n_0 c$. The domain size is $L_x\times L_z\times L_y =1000d_e\times 500d_e \times 500d_e$, where $d_e=c/\omega_{pe}$ is the inertial length. The resolution of the simulations is $N_x\times N_z\times N_y=4096\times2048\times2048$ or $17.2$ billion cells. All simulations used $300$ particles (both species together) per cell with the total number of particles to be $\sim 5.2$ trillion particles. A list of simulation runs and their parameters can be found in Table 1. Simulations employed periodic boundary conditions in the $x$ and $y$-directions, and in the $z$-direction used conducting boundaries for the fields and reflecting boundaries for the particles. A long-wavelength perturbation is included to create a dominating X-line at the center of the simulation domain. Different from the earlier simulations, we also inject an array of perturbations with 
different wavelengths at $t=0$

\begin{eqnarray}
  \delta {\bf B}/B_0&=&\sum_{j,k}a_0\cos(k_j x+k_k
  z+\phi_{j,k})\hat{\bf y}\nonumber\\
  &&+\sum_{l,n}a_0\cos(k_l x +k_n y+\phi_{l,n})\hat{\bf z}
\end{eqnarray}

\noindent with the total summed wave power
$(\delta B/B_0)^2$ up to $0.4$ to initiate a background turbulence at the beginning of the simulation. These initially injected perturbations have $10$ modes with wavelengths
longer than the initial thickness of the current sheet. The simulation lasts $\omega_{pe} t \approx 1100$, which is about the time
for a light wave to travel through $\tau_c = L_x/c$. The timescale to traverse the current sheet
thickness, however, is much shorter. In addition to the antiparallel case ($b_g = 0$), we also included a case with ($b_g = 1$) to examine the influence of a guide field.

In VPIC simulations, we have implemented a particle tracing module to output particle trajectories and find the electric field, magnetic field, and bulk fluid velocity at particle locations \citep{Guo2016,Guo2019,Li2018Role,Li2019Formation,Kilian2020}, and therefore we can evaluate the relative importance of motional electric field $\textbf{E}_m = - \textbf{V} \times \textbf{B}/c$ and non-ideal electric field  $\textbf{E}_n = \textbf{E} + \textbf{V} \times \textbf{B}/c$ based on the generalized Ohm's law \citep{Guo2019}. In this study, we uniformly select one of $50,000$ particles ($\sim100$ million in total) in the beginning of the simulation and analyze their acceleration to high energy. This allows us to quantitatively determine the contribution of Fermi-like acceleration and non-ideal electric field in 3D simulations, respectively.

\begin{table}
\centering
\begin{tabular*}{0.52\textwidth}{ccccc}
\hline 
Run   & $\sigma_e$ & system size & $b_g$ & $\delta B^2/B_0^2$ \\
\hline
3D-1 [S]& 100  &$1000 d_e\times 500 d_e\times 500 d_e$& 0 & 0.1\\ 
3D-2& 400  &$1000 d_e\times 500 d_e\times 500 d_e$&  0 & 0.1\\  
3D-3& 1600 &$1000 d_e\times 500 d_e\times 500 d_e$&  0 & 0.1\\  
3D-4& 25   &$1000 d_e\times 500 d_e\times 500 d_e$&  0 & 0.1\\  
3D-5& 6    &$1000 d_e\times 500 d_e\times 500 d_e$&  0 & 0.1\\  
3D-6& 100 &$1000 d_e\times 500 d_e\times 500 d_e$&  1 & 0.1\\  
3D-7& 100 &$1000 d_e\times 500 d_e\times 500 d_e$&  0 & 0.0\\  
3D-8& 100 &$1000 d_e\times 500 d_e\times 500 d_e$&  0 & 0.2\\  
3D-9& 100 &$1000 d_e\times 500 d_e\times 500 d_e$&  0 & 0.4\\  
3D-10& 100&$1000 d_e\times 500 d_e\times 500 d_e$& 0 & 0.1\\

 \hline
 \end{tabular*}
 \caption{List of simulation runs and their parameters. }
 \label{table1}
\end{table}

\section{Simulation Results} 
\subsection{Plasma Dynamics} 
\subsubsection{General Overview of RTMR} 
\begin{figure}
\centering
\includegraphics[width=0.85\textwidth]{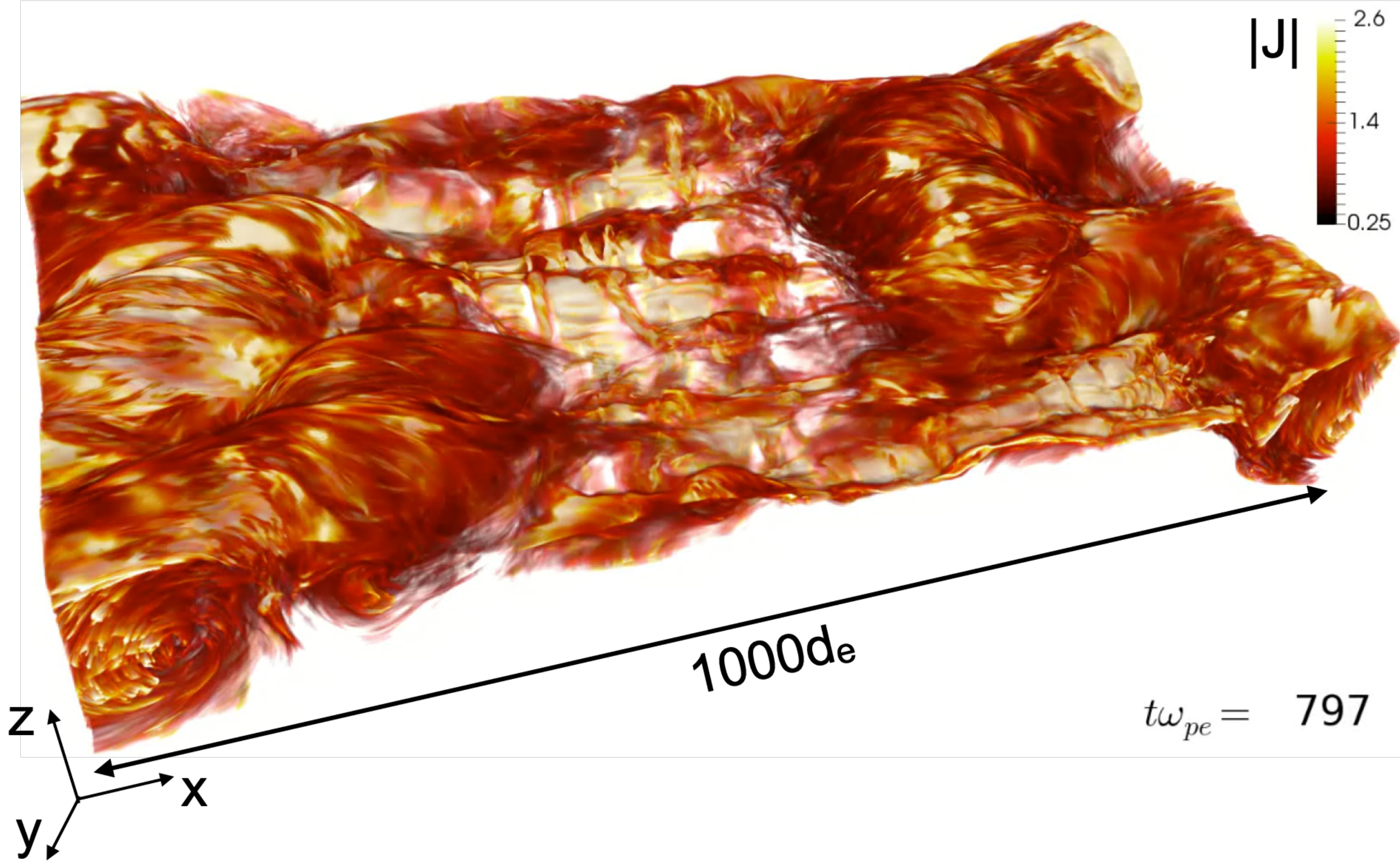}
\caption{\label{fig:jrender}The distribution (volume rendering) of the magnitude of the current density in the reconnection region at $\omega_{pe}t = 797$ for the standard Run 3D-1. Under the influence of injected turbulence, the current sheet breaks into a turbulent reconnection region filled with ample structures such as flux ropes and current sheets.}
\end{figure}

The imposed wave perturbation leads to the evolution of current sheets in a fully 3D fashion. The initial current sheet quickly breaks into a broad reconnection region filled with nonlinear structures such as flux ropes and current sheets, evolving into a RTMR state. Figure~\ref{fig:jrender} shows a volume-rendering diagram of the magnitude of the current density with $\sigma_e = 100$ at $\omega_{pe}t= 797$ (standard Run 3D-1). As reconnection proceeds, anti-parallel magnetic field from upstream continuously feeds into the reconnection region, forming new current sheets at different local regions and the new current sheets further break into small-scale flux ropes (See below for more discussion on flux rope dynamics). This process happens over and over again in a cyclic way. 

\begin{figure}
\centering
\includegraphics[width=0.84\textwidth]{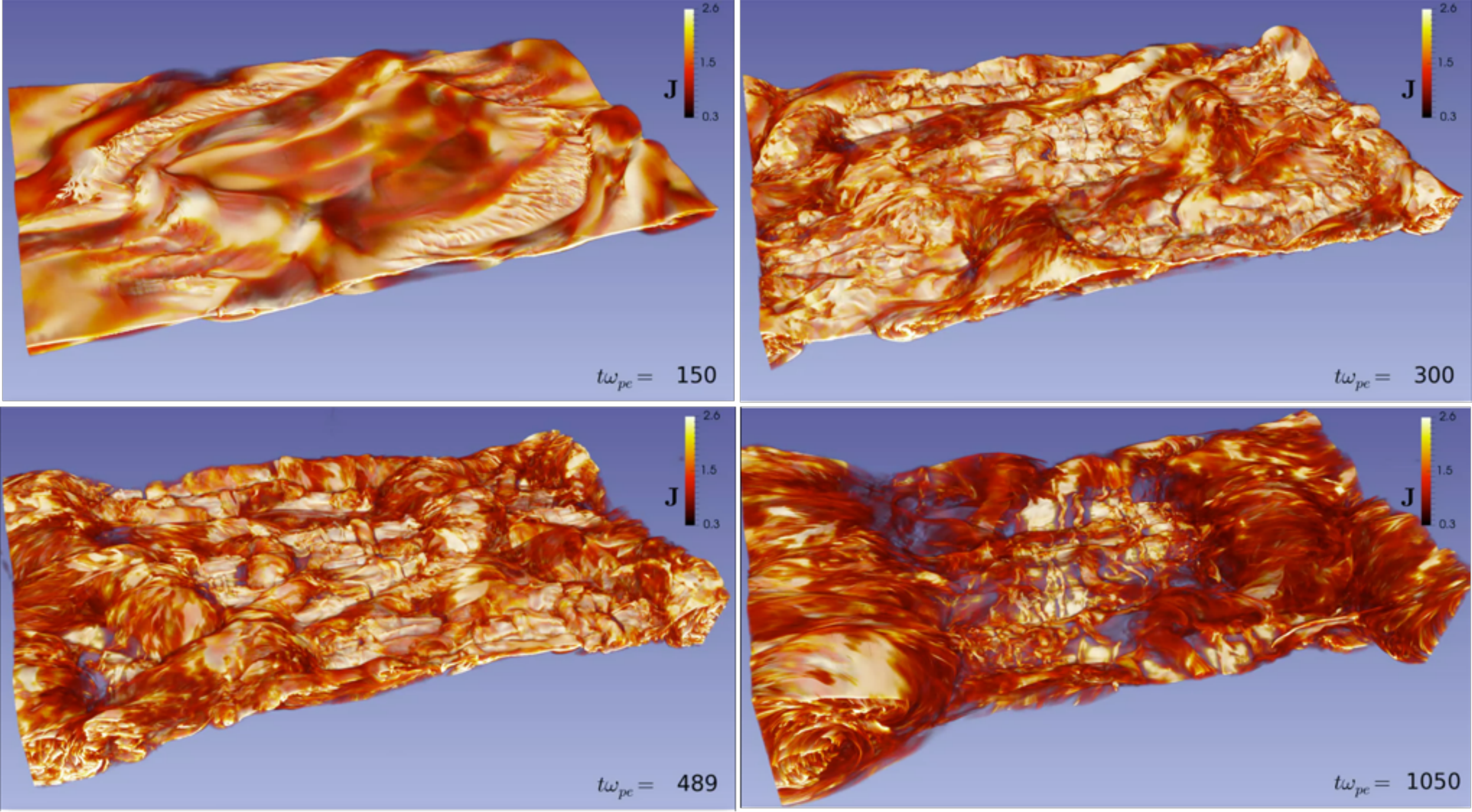}
\caption{\label{fig:jtime}Four snapshots of current density showing the time evolution of the reconnection layer for the Run 3D-1. See also Movie~1 (\url{https://youtu.be/-2EsinquZjA}). Upstream magnetic field continuously feeds into the reconnection region, forming new current sheets and the current sheets keeps breaking into flux ropes.}
\end{figure}

\begin{figure}
\centering
\includegraphics[width=0.84\textwidth]{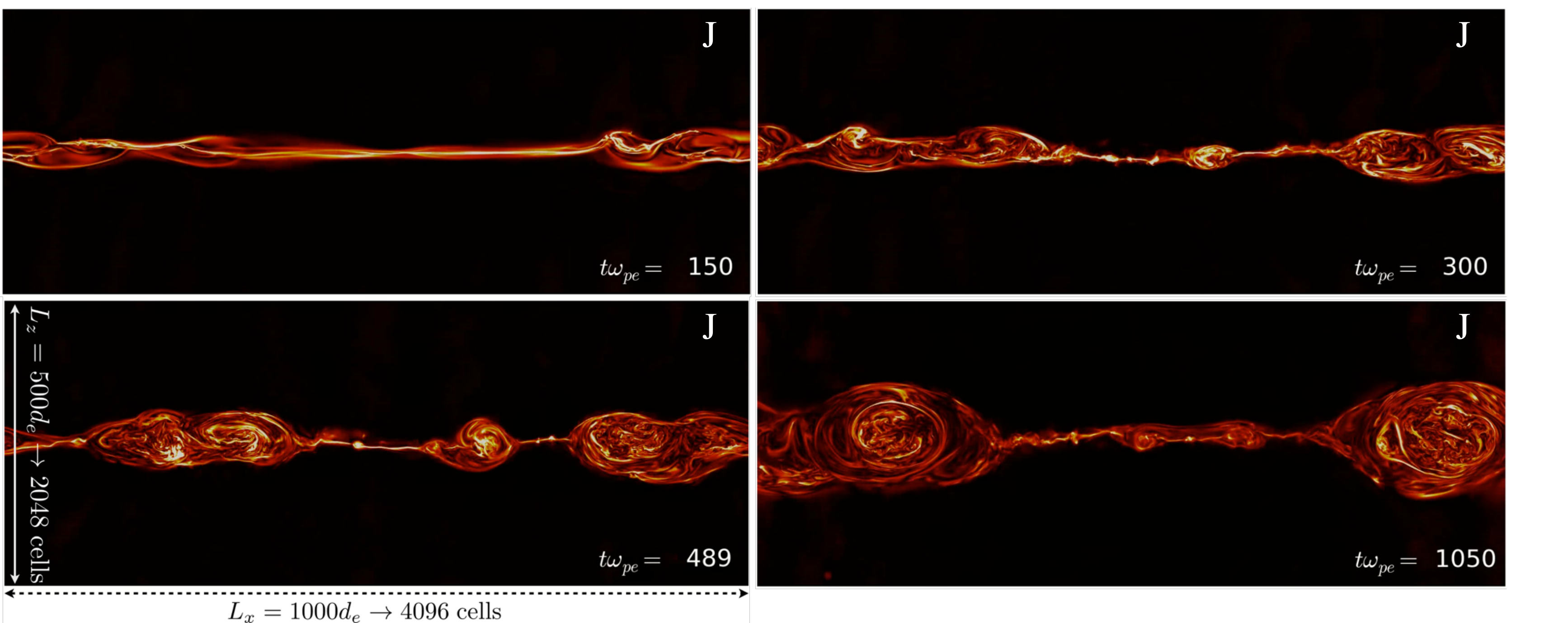}
\caption{\label{fig:jcuts}2D cuts of current density at four different time steps for the Run 3D-1. See also Movie~2 (\url{https://youtu.be/5-eL9oXXCLs}). Kinetic structures with thickness $\sim d_e$ are continuously generated even for the large-scale turbulent reconnection system.}
\end{figure}

Figure~\ref{fig:jtime} provides four additional snapshots to show the time evolution of the reconnection layer (see also Movie~1 \url{https://youtu.be/-2EsinquZjA}). While their 2D counterparts are well studied \citep{Daughton2006,Daughton2007,Liu2015,Liu2020}, the 3D simulations reveal a picture far more complicated and rich in structures.  To see those fine structures more closely, one can review Figure~\ref{fig:jcuts} for four 2D cuts at different time steps and Movie~2 (\url{https://youtu.be/5-eL9oXXCLs}). For the simulations we present here, the kinetic layers $\sim d_e$ are still continuously generated and this feature sustains throughout the dynamical development of reconnection, indicating kinetic effects may still be important for breaking reconnecting field lines even when the system is turbulent and the largest dimension $L_x$ is about a thousand times larger than the kinetic scale. Importantly, such a single $d_e$-scale diffusion region often dominates the primary x-line that separates the reversal outflow jets.  For kinetic simulations we have carried out so far, it appears that the current sheet always collapses to a thin sheet with thickness comparable to kinetic scale in the weak guide field limit.
The cyclic generation of plasmoids is likely related to the loss of current sheet equilibrium due the depleted pressure by reconnection \citep{Liu2020}.

\begin{figure}
\begin{center}
\includegraphics[width=0.6\textwidth]{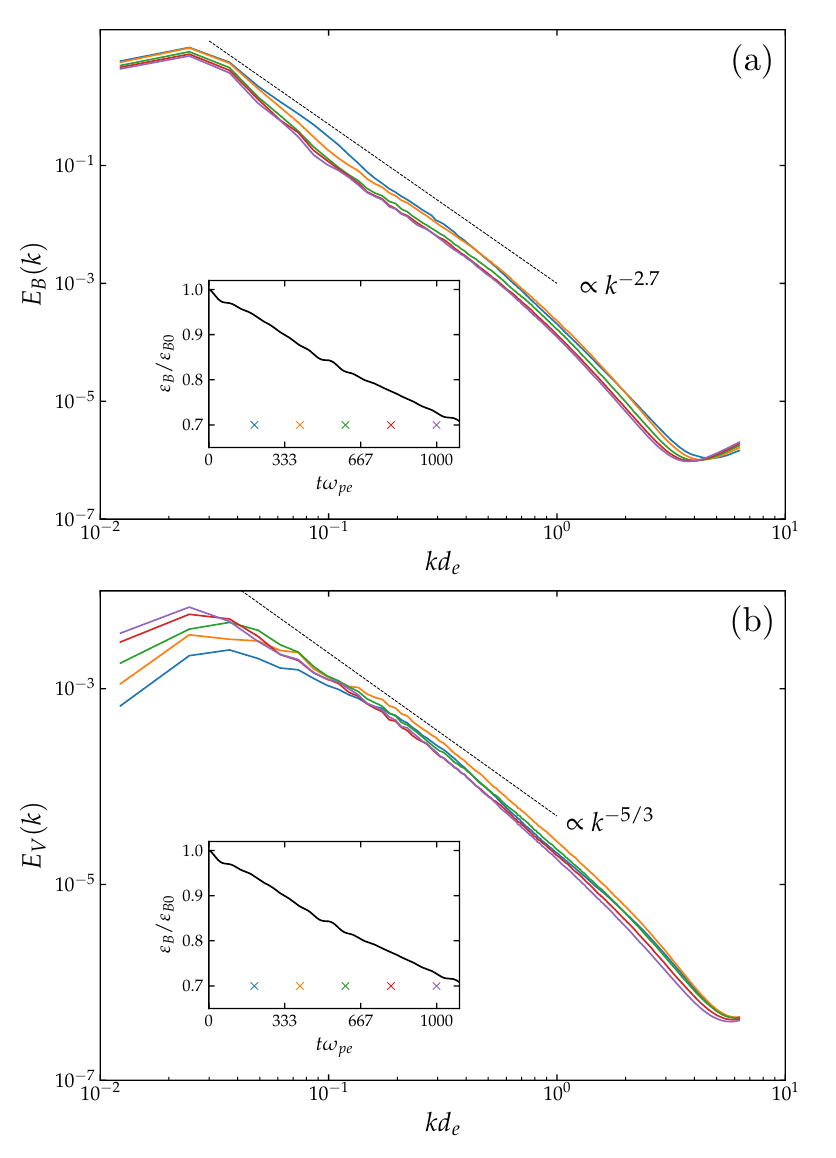}
\caption{\label{fig:power}Magnetic (a) and kinetic (b) power spectra as a function of wave number at different times in the standard Run 3D-1 indicated by ``$\times$'' signs in the subpanels. The subpanels also show the evolution of magnetic energy in the system. The magnetic power spectra with scales above the kinetic scale resemble a power-law with a slope about $-2.7$, whereas the kinetic energy spectra agree roughly with a $-5/3$ slope. }
\end{center}
\end{figure}

\begin{figure}
\begin{center}
\includegraphics[width=0.8\textwidth]{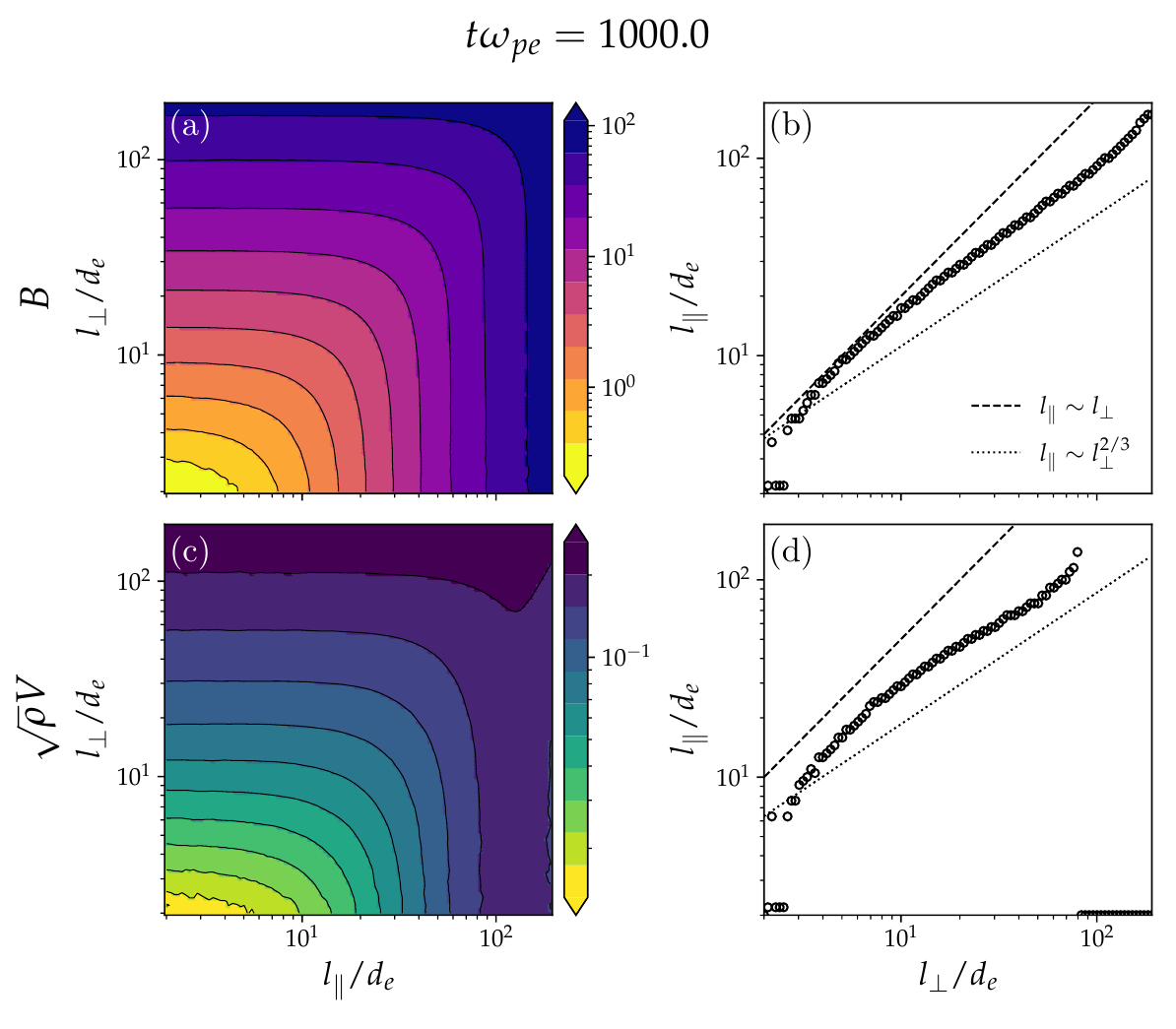}
\caption{\label{Fig:structure-function}Second-order structure functions of the magnetic field and kinetic flow $\vect{w}=\sqrt{\rho}\vect{V}$ in the standard Run 3D-1 (panel a and c), following the method by~\citet{Cho2000}. In the method, the local magnetic fields are calculated as $\vect{B}_l=(\vect{B}(\vect{r}_1) + \vect{B}(\vect{r}_2))/2$, and the second-order structure functions for $\vect{B}$ and $\vect{w}$ are $F_2^b(l_\parallel, l_\perp)=\left<|\vect{B}(\vect{r}_1) - \vect{B}(\vect{r}_2)|^2\right>$ and $F_2^w(l_\parallel, l_\perp)=\left<|\vect{w}(\vect{r}_1) - \vect{w}(\vect{r}_2)|^2\right>$, respectively. $l_\parallel=\vect{l}\cdot\vect{B}_l/|\vect{B}_l|$ and $l_\perp=\sqrt{l^2-l_\parallel^2}$, where $\vect{l}=\vect{r}_2-\vect{r}_1$ and $l=|\vect{l}|$.
Panels (b) and (d) represent the relationships between parallel scale $l_\parallel$ and perpendicular scale $l_\perp$ of the contours
in panels (a) and (c), measuring the scale dependency of turbulent eddy anisotropy. The dashed and dotted lines represent the relations $l_\parallel\sim l_\perp$ (scale-independent) and $l_\parallel\sim l_\perp^{2/3}$ (GS theory), respectively.}
\end{center}
\end{figure}

To reveal the turbulent nature of the reconnection layer, we plot the magnetic and kinetic power spectra at different simulation times in Figure \ref{fig:power}. The embedded figures are the evolution of magnetic field energy with ``$\times$'' signs indicating times when those power spectra are measured. The magnetic power spectrum above kinetic scale $k d_e < 1$ resembles a power law ``inertial'' range with slope about $\sim -2.7$, decaying in general as magnetic field energy is converted into particle energy. Meanwhile, the kinetic energy spectra are more or less in consistency with the classic -5/3 spectrum. More detailed analysis (not shown) suggests that the magnetic power and kinetic power with wave number vector in the reconnection plane ($x-z$) are much stronger than along the current sheet direction (the $y$ direction). The slope of the magnetic power spectra appears to be strongly mediated by reconnection, likely due to different processes such as forward cascade and inverse cascade through secondary instabilities and flux rope merging \citep{Bowers2007,Daughton2014,Guo2015,Huang2016,Loureiro2018,Yang2020,Zhou2020,Kowal2020}. However, we acknowledge that this issue is still under debate. For example, \citet{Huang2016} present resistive MHD simulations with high Lundquist number showing that several properties of reconnection driven turbulence, such as the magnetic energy spectrum and anisotropy clearly deviated from the classical Goldreich-Sridhar model. However, \citet{Kowal2017} presented simulations over a much longer time scale, and they showed that the slope gradually evolves toward $5/3$. Due to the added on diagnostics for particle acceleration, our simulations only extend to a limit time. We defer more studies targeting on this issue to a future publication\footnote{Note that the often used periodic boundary conditions may not be appropriate for studying the long-term evolution of the system beyond several light crossing times $\tau_c$. Our 3D simulations with reduced domain sizes (not shown) suggest that when the simulations extend to $\sim 10 \tau_c$, the central reconnection layer becomes saturated with no net outflow or inflow, while some local patches of reconnection still occurs in the 3D simulations and consumes the upstream magnetic field.}.
We further look into the structure functions and the scale dependent turbulent eddy anisotropy in the reconnection generated turbulence in Figure \ref{Fig:structure-function}. The contours of the structure fucntions shown in Figure \ref{Fig:structure-function} (a) and (c) clearly show elongated turbulent eddies along the local magnetic field direction. The turbulence anisotropy scaling (Figure \ref{Fig:structure-function} (b) and (d)) shows that the turbulence is scale-independent at small scales ($l_\parallel< 20d_e$) and then resembles Goldreich-Sridhar (GS) scaling $l_\parallel \sim l_\perp^{2/3}$~\citep{Goldreich1995,Goldreich1997} at larger scales, similar to the MHD simulation results by \citet{Kowal2017}.

\subsubsection{Dynamical Evolution of Flux Ropes} 
\begin{figure}
\centering
\includegraphics[width=0.9\textwidth]{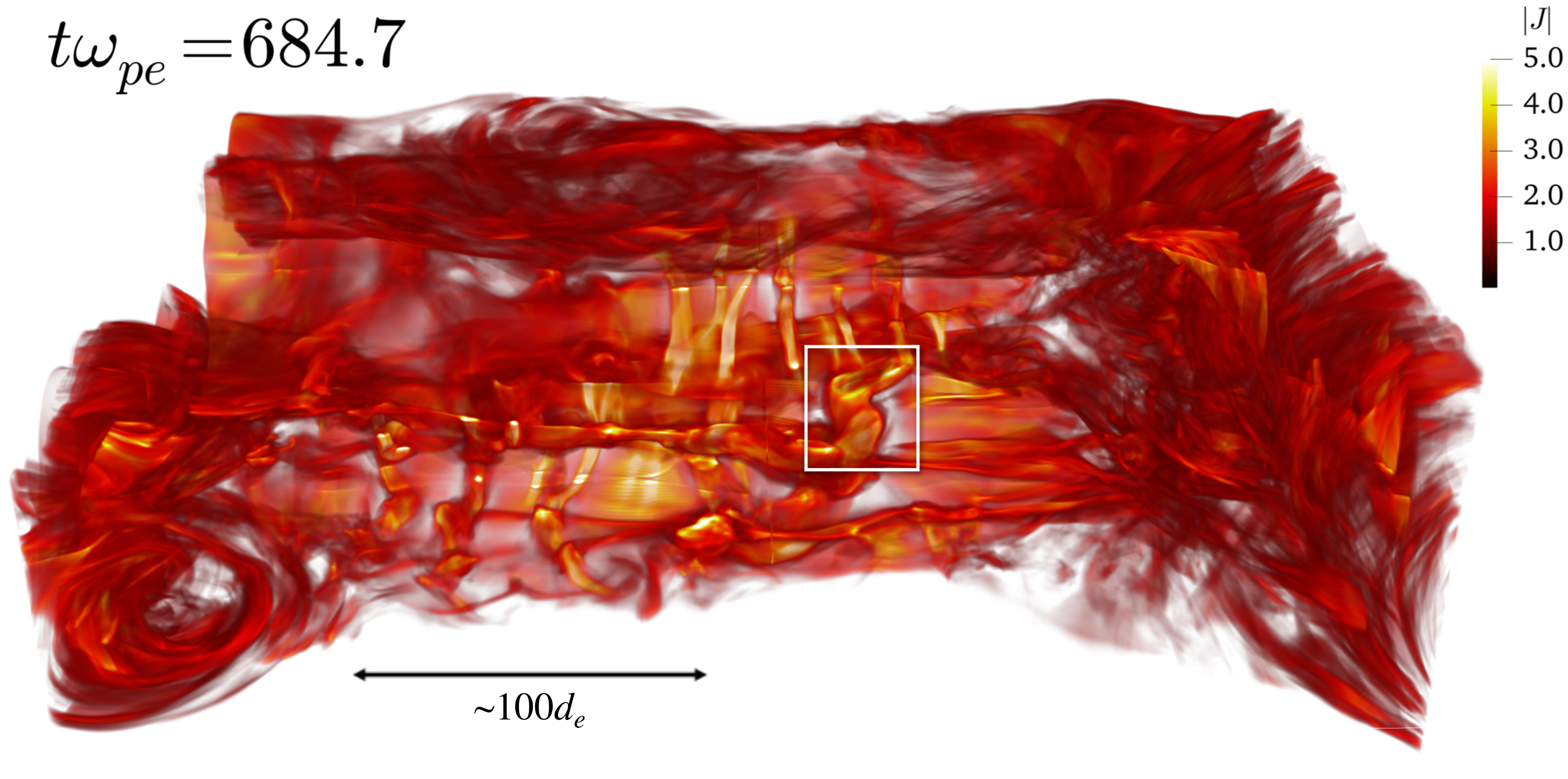}
\caption{\label{fig:fluxropes} Volume rendering of magnitude of current density for a selected region ($250<x/d_e<680$ and $250<y/d_e<500$) in the reconnection layer where multiple flux ropes are generated during magnetic reconnection. See also Movie~3 (\url{https://youtu.be/FWI-Fhvgsrc}). A flux rope is labeled using a white window for further study in Figure~\ref{fig:single_FR}.}
\end{figure}

\begin{figure}
\centering

\includegraphics[width=0.9\textwidth]{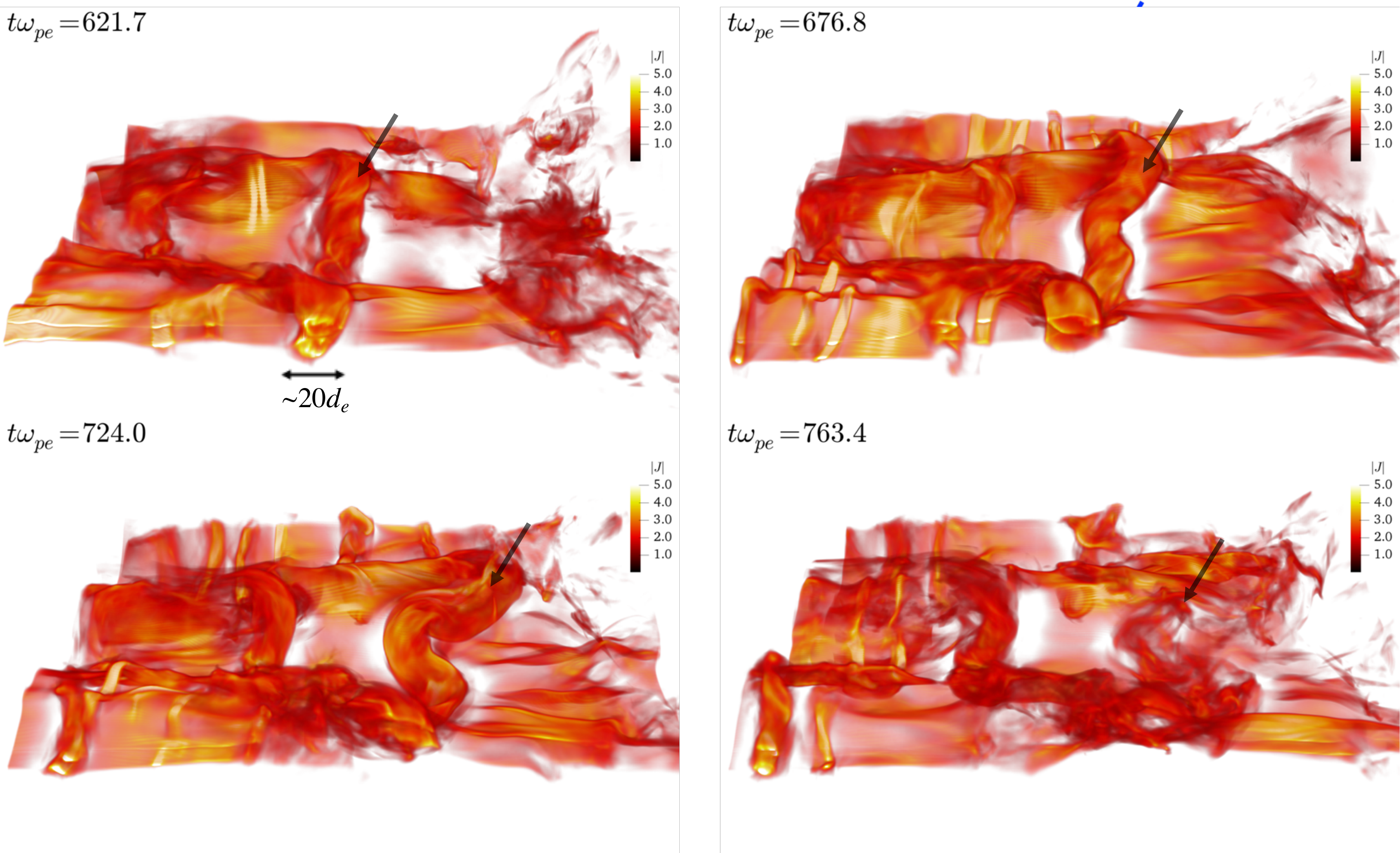}
\caption{\label{fig:single_FR}Several snapshots of current density showing the evolution of a flux rope labeled in Figure~\ref{fig:fluxropes}. See also Movie~4 (\url{https://youtu.be/WXd1kF5wozM}). The flux rope is disrupted quickly after it emerges during the simulation.}
\end{figure}

During RTMR, numerous new, small-scale flux ropes keep emerging from newly formed current layers. Figure~\ref{fig:fluxropes} (see also Movie~3 \url{https://youtu.be/FWI-Fhvgsrc}) shows a part of the simulation domain ($250<x/d_e<680$ and $250<y/d_e<500$) in the reconnection layer where multiple flux ropes  are generated from the layer. This process is sustained throughout the simulation. Each of the flux ropes appears very different from cylindrical magnetic structures indicated by 2D simulations \citep{Guo2015,Sironi2016}. In three dimensions, the flux ropes can be very dynamical and are unstable to the kink instability. Next, we focus on a flux rope labeled by the white window in Figure~\ref{fig:fluxropes}. Figure~\ref{fig:single_FR} and Movie~4 (\url{https://youtu.be/WXd1kF5wozM}) provide a zoom-in view for the evolution of this flux rope. At $\omega_{pe}t = 621.7$, the flux rope (indicated by the arrow) just emerges from the reconnection layer and has a quasi-2D structure. However, it soon starts to twist because of the nonlinear development of the kink instability and possibly additional velocity shear. The whole flux rope breaks as it is strongly distorted, and is eventually dissolved in the reconnection layer. We find that individual flux ropes do interact and merge with each other, evolving into larger flux ropes, but details are much more complicated than their 2D counterparts \citep[e.g.,][]{Guo2015,Sironi2016,Zhang2018}. While recent studies have explored the consequence of radiation signatures of 2D relativistic plasmoid reconnection \citep{Petropoulou2016,Zhang2018}, we show here that when the simulations extend to 3D, the reconnection layer becomes very turbulent and plasma dynamics -- especially the behavior of the flux ropes -- is significantly different from the corresponding structures (plasmoids) in 2D simulations. Exploring 3D effects in those simulations is important to confirm the robustness of previous 2D results.

\subsubsection{Outflow structures} 
\begin{figure}
\centering
\includegraphics[width=\textwidth]{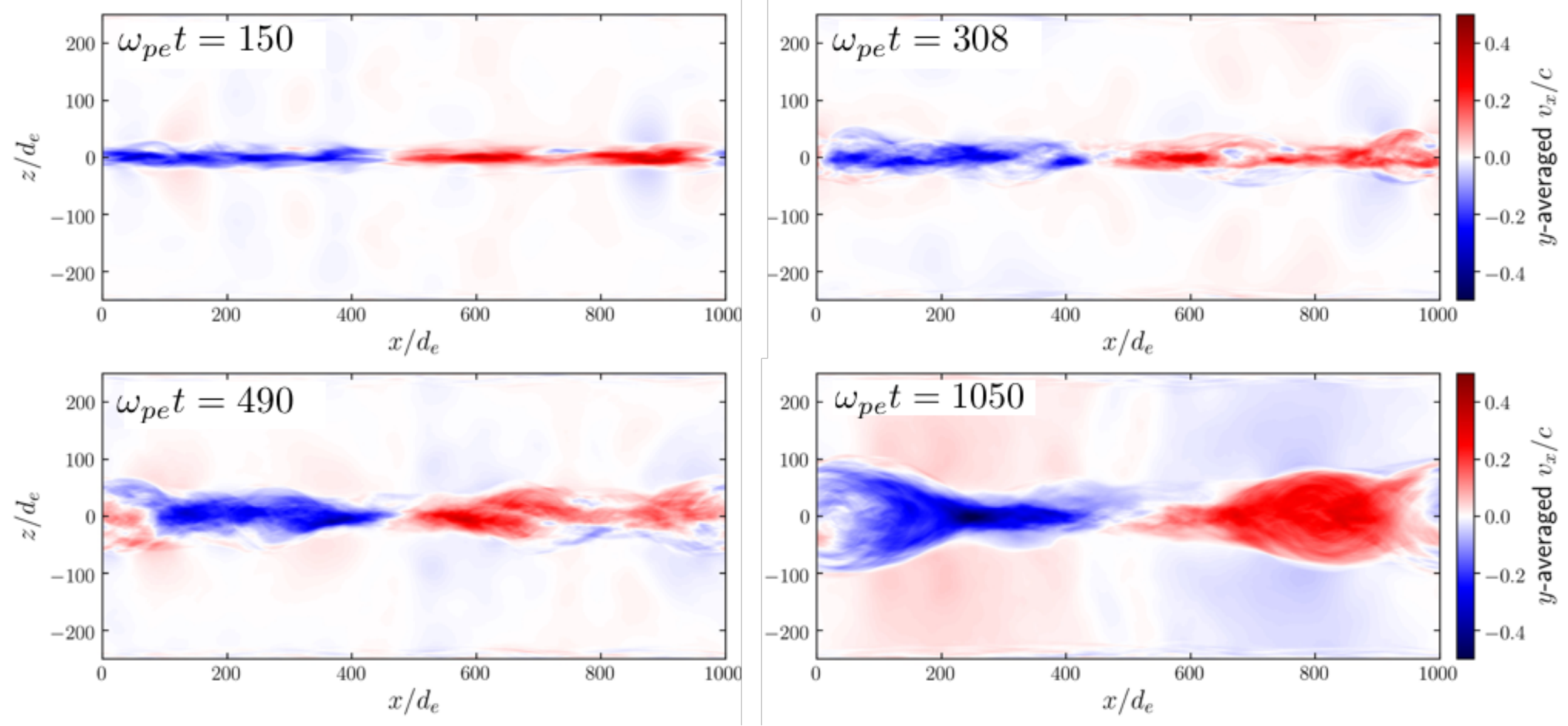}
\caption{\label{fig:vavg}The flow velocity $V_x$ averaged over the $y$ direction. In relativistic turbulent reconnection, the averaged outflow speed is significantly reduced from the theoretical limit $V_A = \sqrt{\sigma/(\sigma+1)}c$ to a fraction of the light speed.}
\end{figure}

\begin{figure}
\centering
\includegraphics[width=0.8\textwidth]{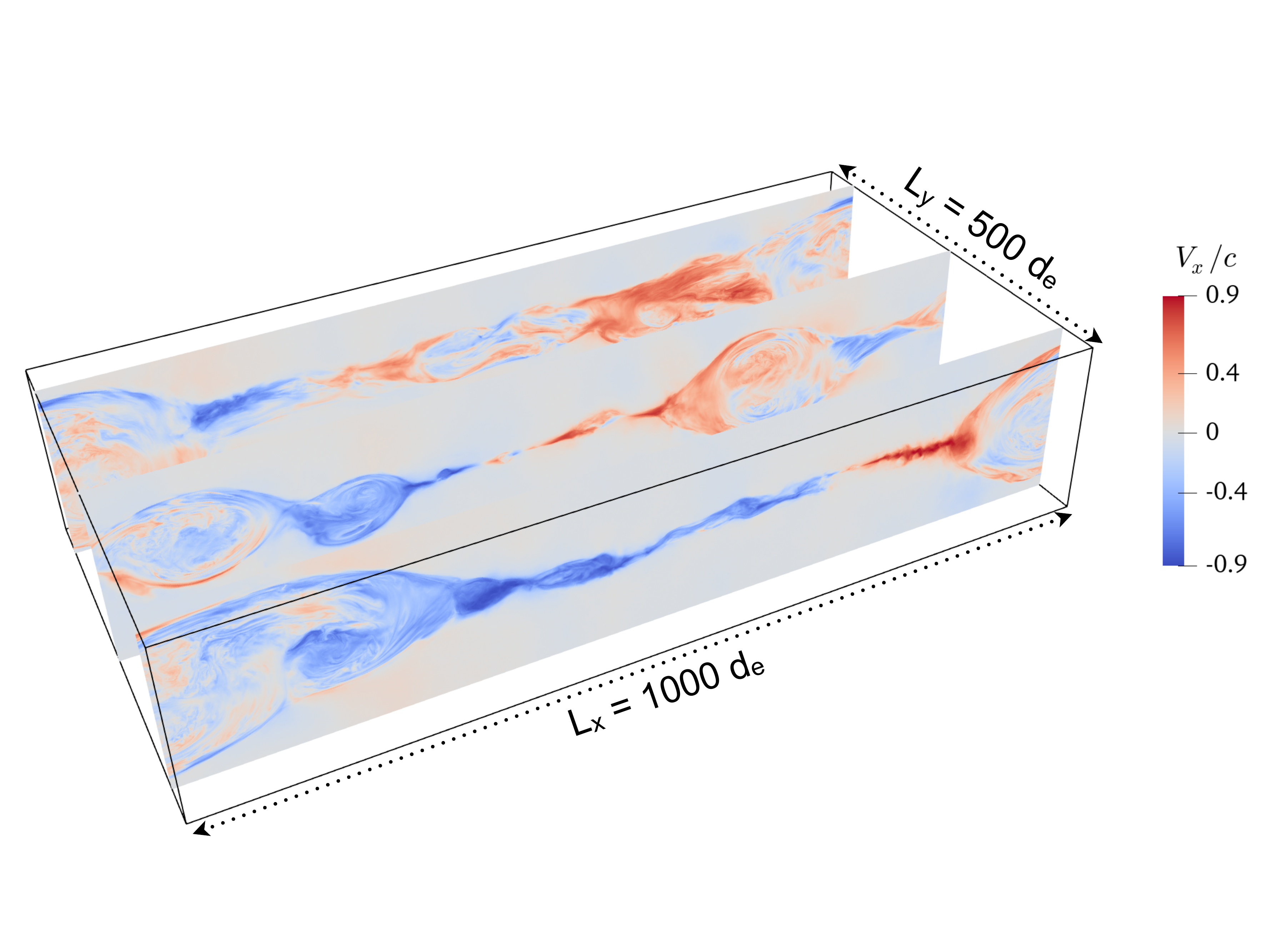}
\caption{\label{fig:vcuts}Three 2D cuts for the outflow speed $V_x$ along different $y$-locations. Because of nonlinear development of instabilities and turbulence, the primary ``x-lines'' in the 2D planes are located differently in x-position along the $y$ axis.}
\end{figure}

It is instructive to explore the reconnection outflow and its structures in the turbulent reconnection layer and how laminar reconnection is modified by turbulence.  Figure~\ref{fig:vavg} shows the outflow speed $\left<V_x\right>$ averaged over the $y$-direction at several different snapshots. We find that the averaged outflow speed is significantly slower than the theoretical value $V_{out} = \sqrt{\sigma/(1+\sigma)}c$ and earlier reported 2D results \citep{Guo2015,Sironi2016}. Instead, the averaged speed in the reconnection layer can only reach $\sim 0.4 c$. The whole reconnection layer appears to be broadened although kinetic layers still develop locally (Figure~\ref{fig:jcuts}). A similar analysis that used resistive MHD simulations  has seen similar effects \citep{Huang2016,Kowal2017}. Figure~\ref{fig:vcuts} shows three different $x-z$ cuts at different $y$ positions at time $\omega_{pe}t = 797$.  The primary x-lines at different y-position that emits reversal jets appear to be much thinner (close to the kinetic-scale). However, the main x-line location in the reconnection plane varies along the $y$-direction, leading to a slower outflow when averaged over a finite distance. This is because the nonlinear development of the kink mode disturbs x-lines and the x-point region so they cannot align along the y-direction as in 2D. When averaged out in 3D, the island structure is no longer significant.

\subsection{Particle Acceleration and its Mechanism} 
\subsubsection{Energy Spectrum}
\begin{figure}
\centering
\includegraphics[width=0.7\textwidth]{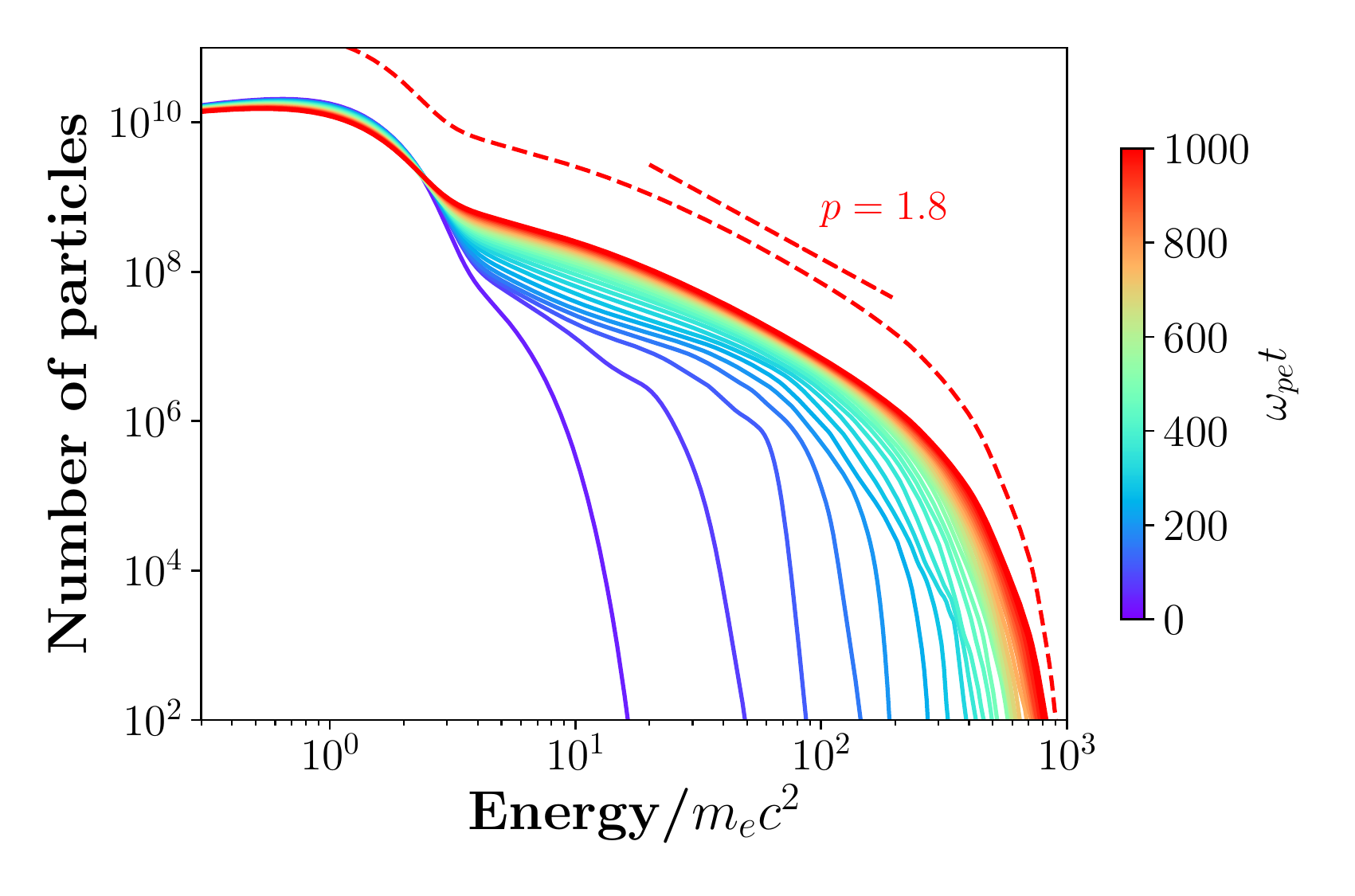}
\caption{\label{fig:spect_time}Particle energy spectrum at different simulation time for the standard run. The spectrum at the last time step is re-plotted and shifted up by a factor of $10$. The spectral index is $p=1.8$.}
\end{figure}

\begin{figure}
\centering           
\includegraphics[width=0.6\textwidth]{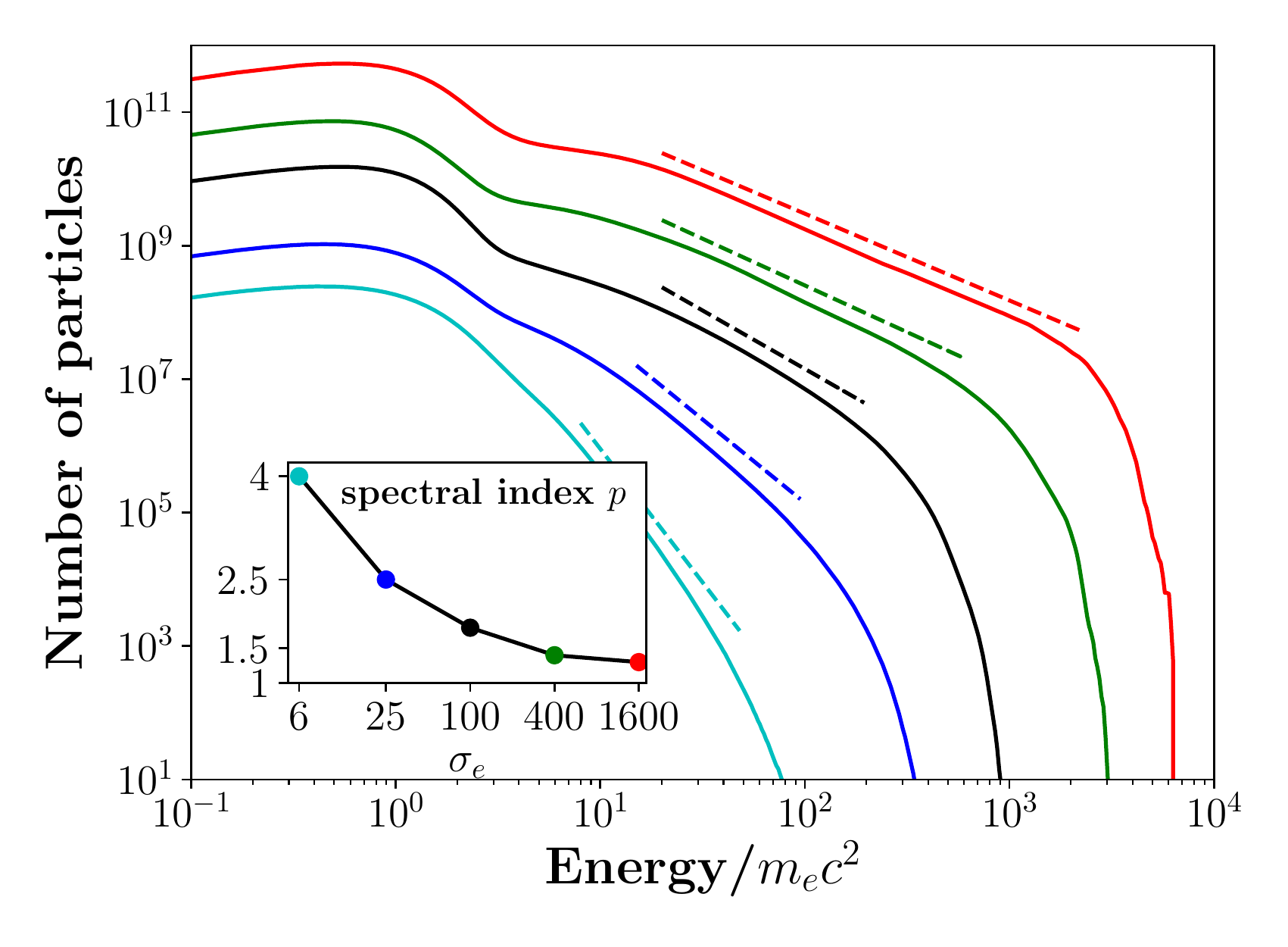}
\caption{\label{fig:spect_sigma}Particle energy spectra at the end of the simulations for different 3D runs with injected turbulence amplitude $\delta B^2/B_0^2 = 0.1$ and different $\sigma_e = 6 - 1600$ (shifted vertically to show the difference). The subpanel shows spectral index for each energy spectrum.}
\end{figure}

Figure~\ref{fig:spect_time} shows the time evolution of particle energy spectra integrated over the whole simulation for $\sigma_e = 100$ (Run 3D-1). The resulting energy spectrum eventually resembles a power-law distribution $f \propto (\gamma - 1)^{-p}$ with $p \sim 1.8$. This is similar to -- but somewhat softer than -- earlier results from 2D and 3D simulations starting from a laminar layer \citep{Guo2014,Guo2015}. Figure~\ref{fig:spect_sigma} shows the energy spectra for a number of 3D runs from $\sigma_e = 6$ to $1600$ with $\delta B^2/B_0 = 0.1$. We observe clear signatures of nonthermal power-law distributions when $\sigma_e > 1$. The spectral index changes from $p = 4$ for $\sigma_e = 6$ to $p = 1.3$ for $\sigma_e = 1600$. The break energy is roughly a few times $\sigma_e$. In general, these results are aligned with 2D simulations and 3D simulations without initial turbulence \citep{Sironi2014,Guo2014,Guo2015,Werner2016}. While the reconnection X-points are strongly modified by the injected fluctuations and self-generated fluctuation due to secondary tearing and kink instabilities, the acceleration of high energy particles is robust and not strongly dependent on the injected turbulence (see more discussion below). This suggests that X-point acceleration is not essential for particle acceleration in forming power-law distributions, as concluded by \citet{Guo2019} and 3D nonrelativistic studies \citep{Dahlin2017,Li2019Formation,Zhang2021b}. Nevertheless, these results suggest that relativistic magnetic reconnection is a robust mechanism for producing nonthermal particle acceleration even in the presence of large-amplitude turbulence.

\subsubsection{Acceleration Mechanism}
\begin{figure}
\centering
\includegraphics[width=\textwidth]{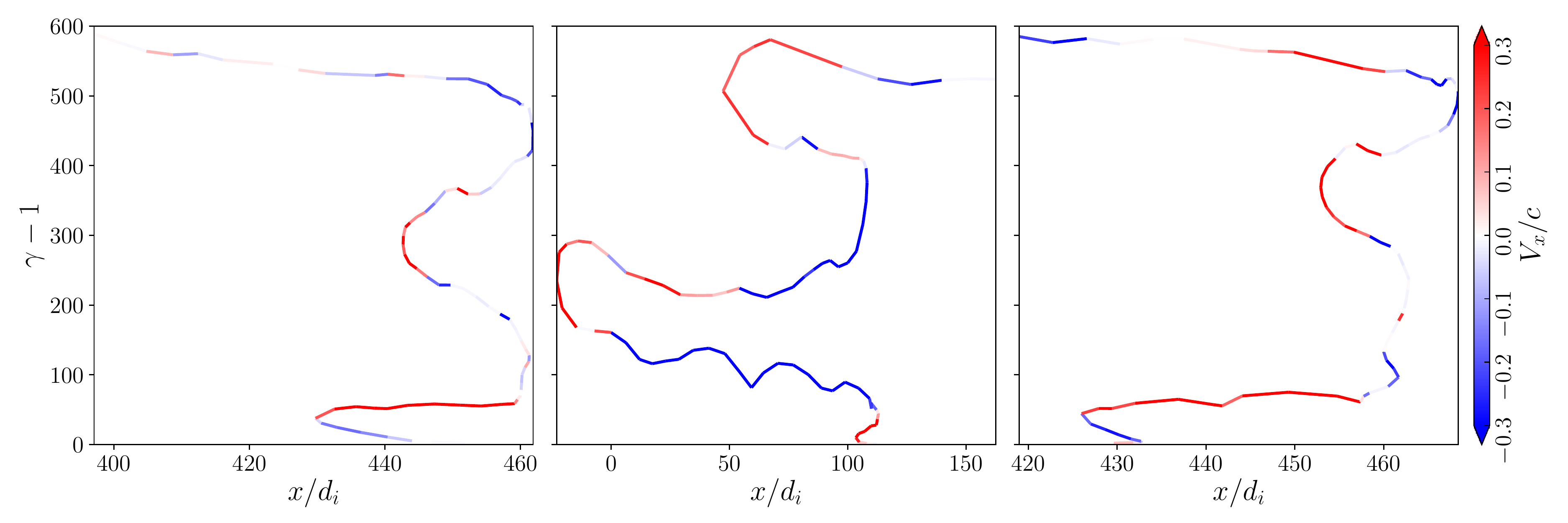}
\caption{\label{fig:traj}Several particle trajectories showing energy versus $x$ position. The color on the curve represents fluid velocity at the $x$ direction $V_x$. These clearly show that Fermi bounces still exist in 3D turbulent reconnection.}
\end{figure}

We now discuss the acceleration of energetic particles to high energy during RTMR. Figure~\ref{fig:traj} shows several sample trajectories presented as particle energy ($\gamma - 1$) versus $x$. These particles are representative ones that are accelerated to very high energy with $\gamma$ reaching several hundred. The color of the curves represents the flow velocity at the particle location in the $x$ direction $V_x$. These particle trajectories clearly show Fermi bounces and during that particles gain a significant amount of energy when they have head-on ``collisions'' with the reconnection generated flows (due to either outflow in the exhaust region or flux rope motions). We have examined hundreds of trajectories and find that this Fermi acceleration process is the main acceleration mechanism for particles accelerated to high energy. While earlier studies have included particles accelerated in simulations with three dimensions, most of analysis on particle acceleration are still based on 2D simulations \citep[e.g.,][]{Guo2014,Guo2015,Guo2019}. Therefore, it is important to confirm that Fermi acceleration is still the dominant acceleration process in 3D RTMR simulations.

\begin{figure}
\centering
\includegraphics[width=0.6\textwidth]{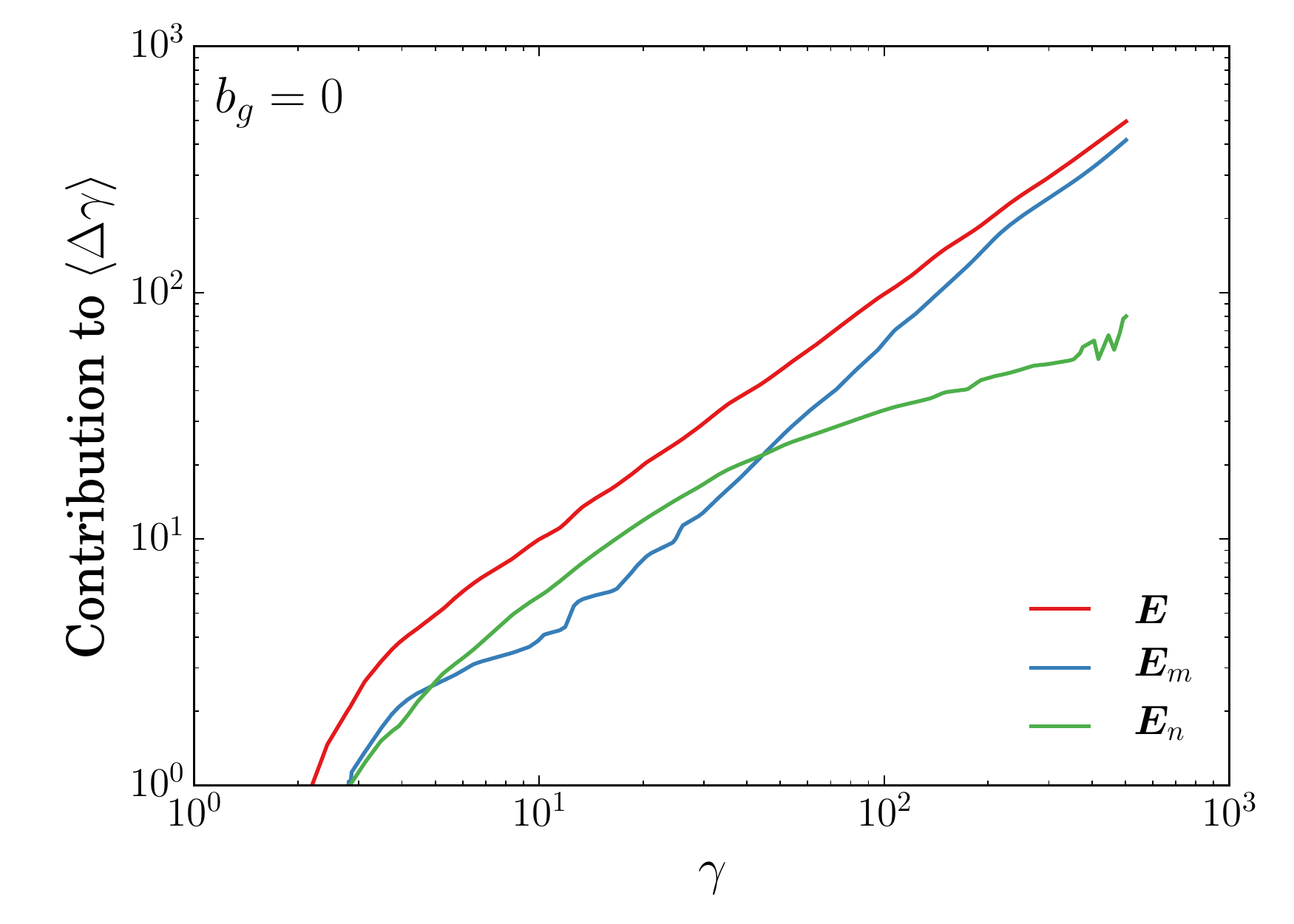}
\caption{\label{fig:egain}The averaged contribution of the motional electric field $\boldsymbol{E}_m = - \boldsymbol{V} \times \textbf{B} /c$ versus the that of the nonideal electric field $\boldsymbol{E}_n=\boldsymbol{E}-\boldsymbol{E}_m$ to the total energy gain per particle in the standard run. }
\end{figure}

While analyzing particle trajectories is important for identifying basic acceleration patterns, this has generated significant controversy and confusion in the past, as the presented trajectories are limited to several subjectively hand-selected particles. Therefore, it is important to statistically study the acceleration mechanisms and consider all possibilities without bias.
To show the dominant acceleration mechanism in a statistical way, we have performed another analysis using all $100 $ million tracer particles to calculate the energy gain in the motional electric field $\textbf{E}_m = - \textbf{V} \times \textbf{B} /c$ (supporting Fermi acceleration) and non-ideal electric field $\textbf{E}_n = \textbf{E} - \textbf{E}_m$ (including X-point acceleration) \citep{Guo2019}. Figure~\ref{fig:egain} shows the contribution of $\textbf{E}_m$ and $\textbf{E}_n$ to the averaged total energy gain as a function of particle energy at the end of the simulation.
The figure shows that 
Fermi-like acceleration, which is supported by motional electric field, dominates at high energy. This conclusion is  similar to analysis made in 2D simulations \citep{Guo2014,Guo2015,Guo2019}, indicating that this acceleration process is not strongly modified by 3D effects. It is worth-noting that the difference between particle acceleration in 2D and 3D relativistic reconnection appear to be smaller than their counterparts in the nonrelativistic regime \citep{Dahlin2017,Li2019Formation,Zhang2021b}.

3D RTMR disfavors X-point acceleration, as this process relies on a channel along the non-ideal electric field ($y$-direction in our simulation) with length $L_n >\Delta \gamma m_e c^2 / (q E_n) $, where $\Delta \gamma m_ec^2$ is the amount of energy gain and $E_n \sim 0.1 B_0$ \citep[see][and discussion in Section 3.3]{Liu2017}. For $\Delta \gamma = 100$ in the case with $\sigma_e = 100$, $L_n$ has to be at least $100 d_e$. This can hardly be satisfied because the kinked flux ropes perturb X-lines so the X-line does not form a potential channel along the y-direction, as we have shown in Figure~\ref{fig:vcuts}.
In addition, 3D simulations have shown that the 3D structure of the parallel electric field is patchy \citep{Karamabadi2013}, indicating it is difficult for it to accelerate particle coherently, except in the beginning of the simulation. Although Fermi acceleration also relies on the $E_y$, the motional electric field is usually $5-10$ times larger than the non-ideal electric field~\citep{Guo2019}, and thus can accelerate particles in a much shorter distance. This is why we still observe clear Fermi bounces in RTMR simulations. The fact that particle acceleration results do not strongly depend on 3D effects favor Fermi acceleration and disfavor X-point acceleration as a main mechanism responsible for producing nonthermal particles in the reconnection region.

\begin{figure}
\centering
\includegraphics[width=0.6\textwidth]{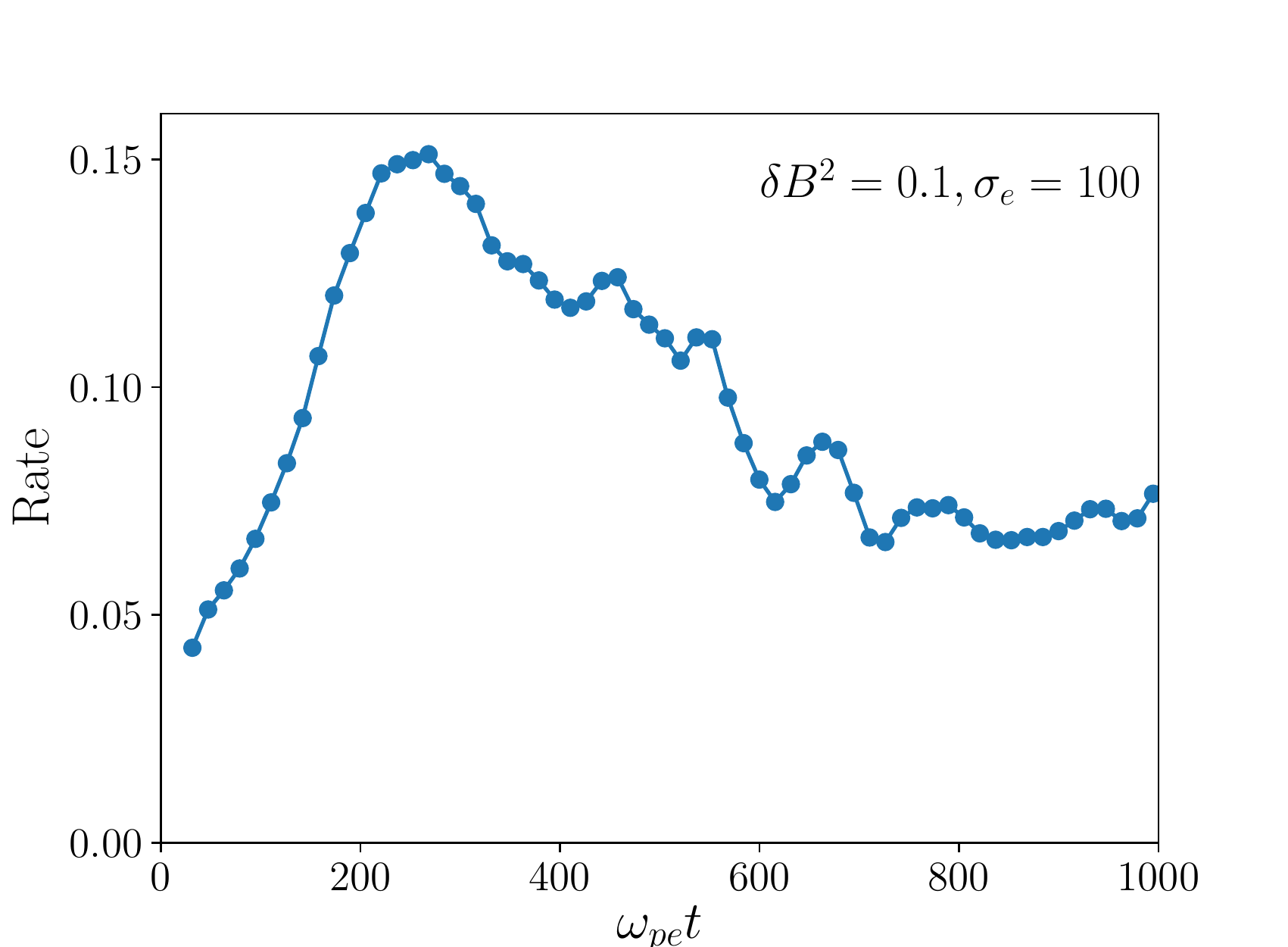}
\caption{\label{fig:rrate}Time evolution of the reconnection rate $R =E_\text{rec}/B_0$ run 3D-1, where $E_\text{rec}$ is the reconnection electric field. The peak rate is about $0.15$, consistent with 2D simulations despite the existence of turbulence.}
\end{figure}

\begin{figure}
\centering
\includegraphics[width=0.6\textwidth]{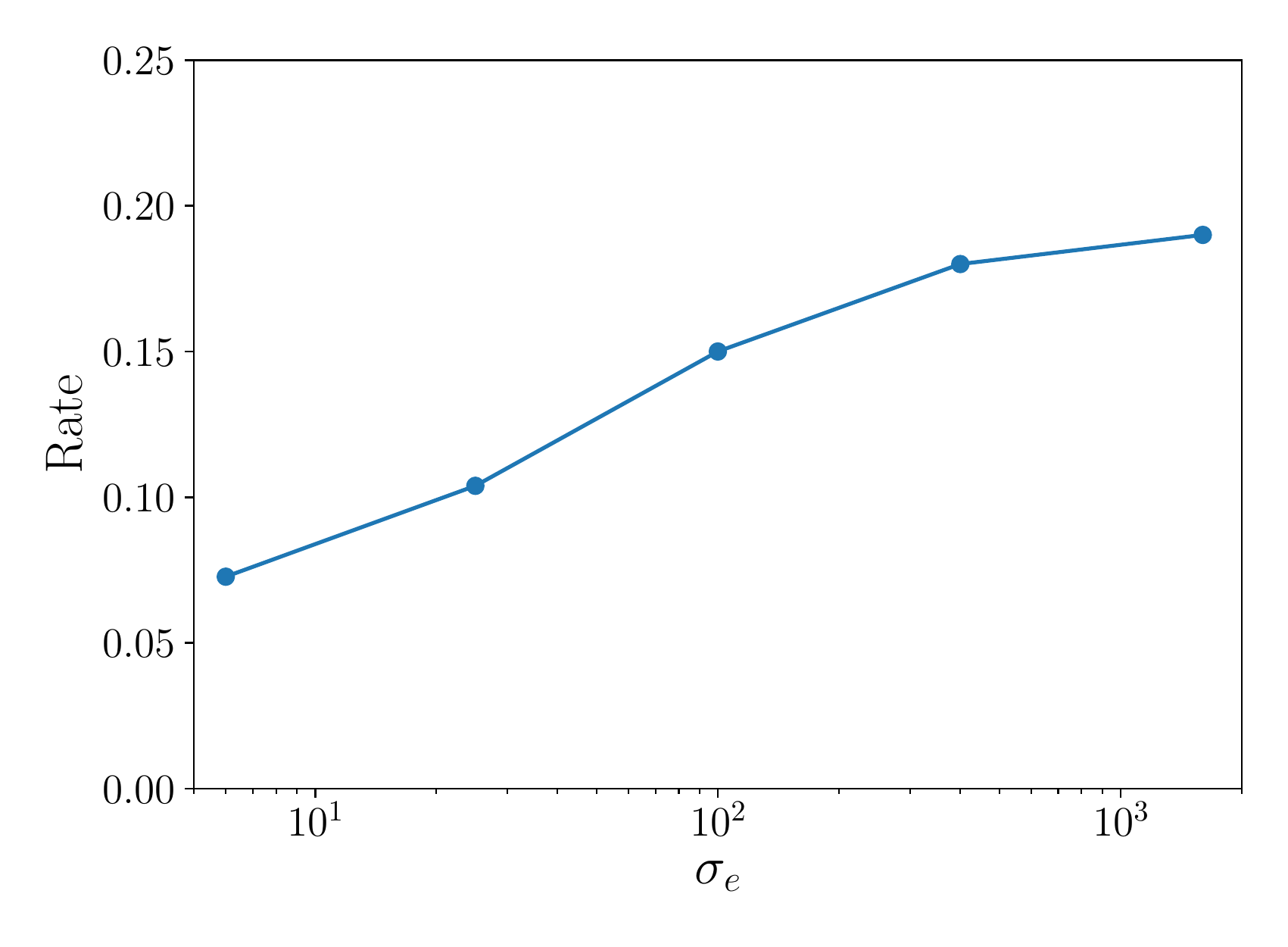}
\caption{\label{fig:rrate-sigma} The peak reconnection rate as a function of $\sigma_e$. The peak rate increases with $\sigma_e$.}
\end{figure}

\subsection{Reconnection Rate}
The reconnection rate indicates the time scales for magnetic reconnection to dissipate magnetic field and the magnitude of electric field that accelerates particles to high energy. Figure~\ref{fig:rrate} shows the time-dependent reconnection rate normalized using the initial asymptotic magnetic field $R = E_\text{rec}/B_0$ in Run 3D-1. This is measured using a technique that determines the separatrices by finding the plasma mixing boundary of upper box ($z>0$) and lower box ($z<0$) particles, described by \citet{Daughton2014}. Because of the initial driving, the current layer quickly becomes unstable, as shown in Figure~\ref{fig:jtime}, and particles from the upper and lower half box can efficiently mix with each other through following the turbulent magnetic field lines (see more discussions in Section~\ref{sec:superdiff}). As a result, reconnection quickly onsets and starts to convert energy violently and reconnection rate starts to be fast $R \sim 0.1$.
The rate peaks around $R = 0.15$ and gradually saturates to about 0.07. Figure~\ref{fig:rrate-sigma} shows the peak reconnection rate for different $\sigma_e$. The peak rate increases with $\sigma_e$ and may saturate around $R \sim 0.2$. This result is in general consistent with results from earlier 2D and 3D reconnection simulations in the relativistic regime \citep{Guo2015,Liu2015,Liu2017} and simulations in the nonrelativistic regime \citep{Liu2017,Daughton2014}, suggesting that the injected and self-generated turbulence cannot enhance the reconnection rate in kinetic simulations. This may contradict the predictions by turbulent reconnection models~\citep{Lazarian1999} but is understandable because the reconnection rate in kinetic simulations without driving already approaches the theoretical upper limit~\citep{Liu2017}. The consistent result across different simulations suggests a universal value for magnetic reconnection rate in the magnetically dominated regime, at least for reconnection starting from a long current sheet with thickness of tens of skin depths.

\begin{figure}
\centering
\includegraphics[width=0.6\textwidth]{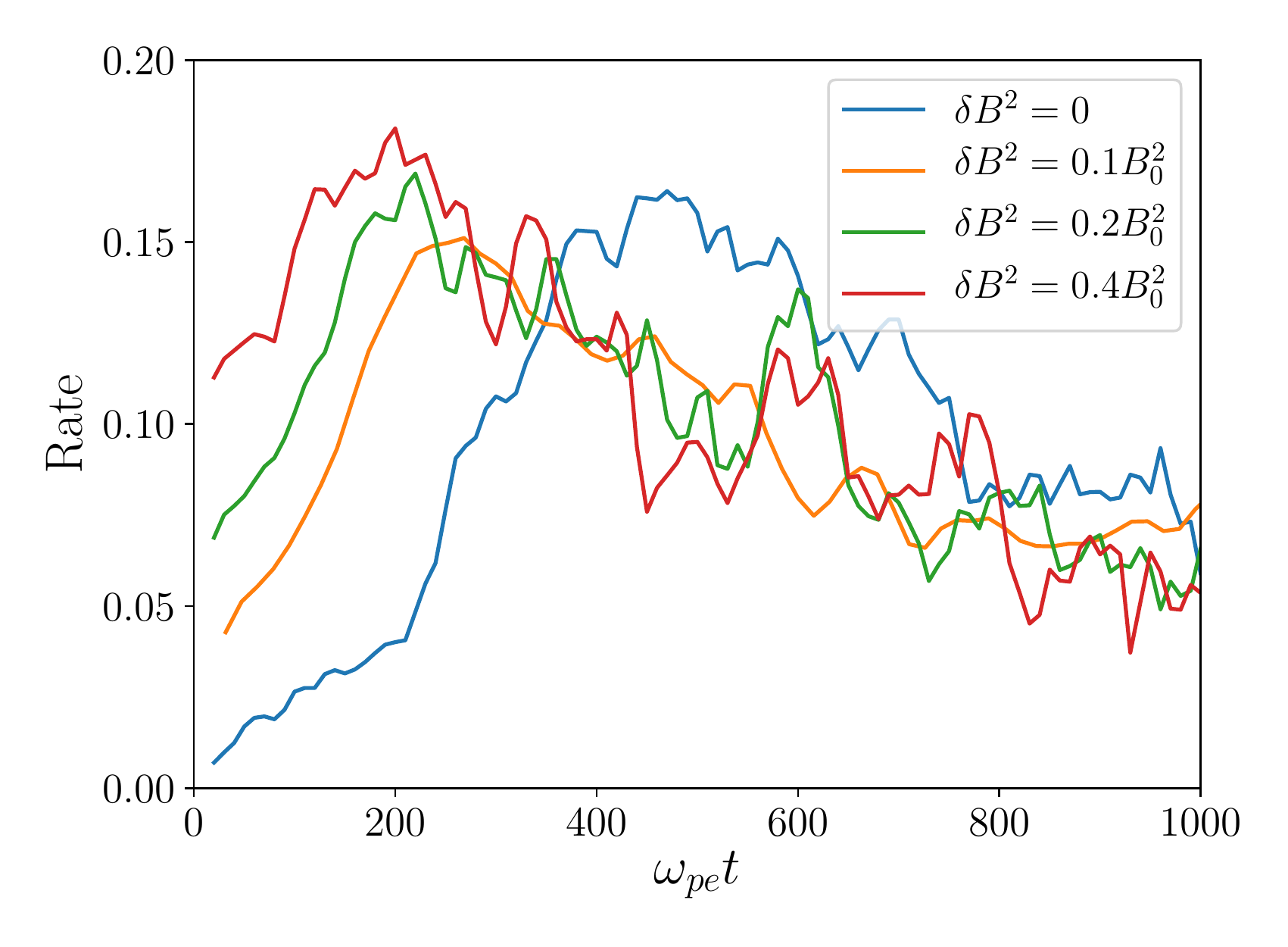}
\caption{\label{fig:rate_deltab}The reconnection rate for the runs with different turbulence amplitude.  The injected fluctuation leads to a shorter onset time where reconnection becomes fast $R \sim 0.1$. The peak reconnection rate does not strongly depend on the turbulence amplitude.}
\end{figure}

\subsection{Dependence on Turbulence Amplitude}
\label{sec:turb_amp}

In existing theories and MHD simulations of turbulent reconnection, the turbulence amplitude strongly influences the reconnection rate and particle acceleration \citep{Lazarian1999,Kowal2009}.  Here we use a series of kinetic simulations with initial driving to examine the role of initial turbulence amplitude on reconnection rate. With the same set of runs, we also study how initial turbulence affects particle acceleration.

Figure~\ref{fig:rate_deltab} compares reconnection rate as a function of time for different initial perturbation amplitude $\delta B^2 / B_0^2= 0, 0.1, 0.2$ and $0.4$. We note that even for the case with $\delta B^2 = 0$, there is still substantial self-generated fluctuations because of secondary tearing and kink instability, similar to earlier studies~\citep{Guo2014,Guo2015}. Figure~\ref{fig:rate_deltab} shows that the peak reconnection rate does not strongly depend on the turbulence level. This suggests that for the regime we explore, the peak reconnection rate may still be determined by kinetic physics, as indicated in Figure~\ref{fig:jcuts}. This result is also in contrast with recent high Lundquist number MHD simulations \citep{Yang2020}, which shows that the reconnection rate is around $R \sim 0.01$ without turbulence injection and the injected turbulence does enhance the reconnection rate to $R \sim 0.1$. Although the presence of initial turbulence does not change the peak reconnection rate, simulations with higher initial driving onset and achieve peak reconnection rate faster. This would indicate the turbulence can accelerate the ``triggering'' process even if the rate does not change much.

\begin{figure}
\centering
\includegraphics[width=0.6\textwidth]{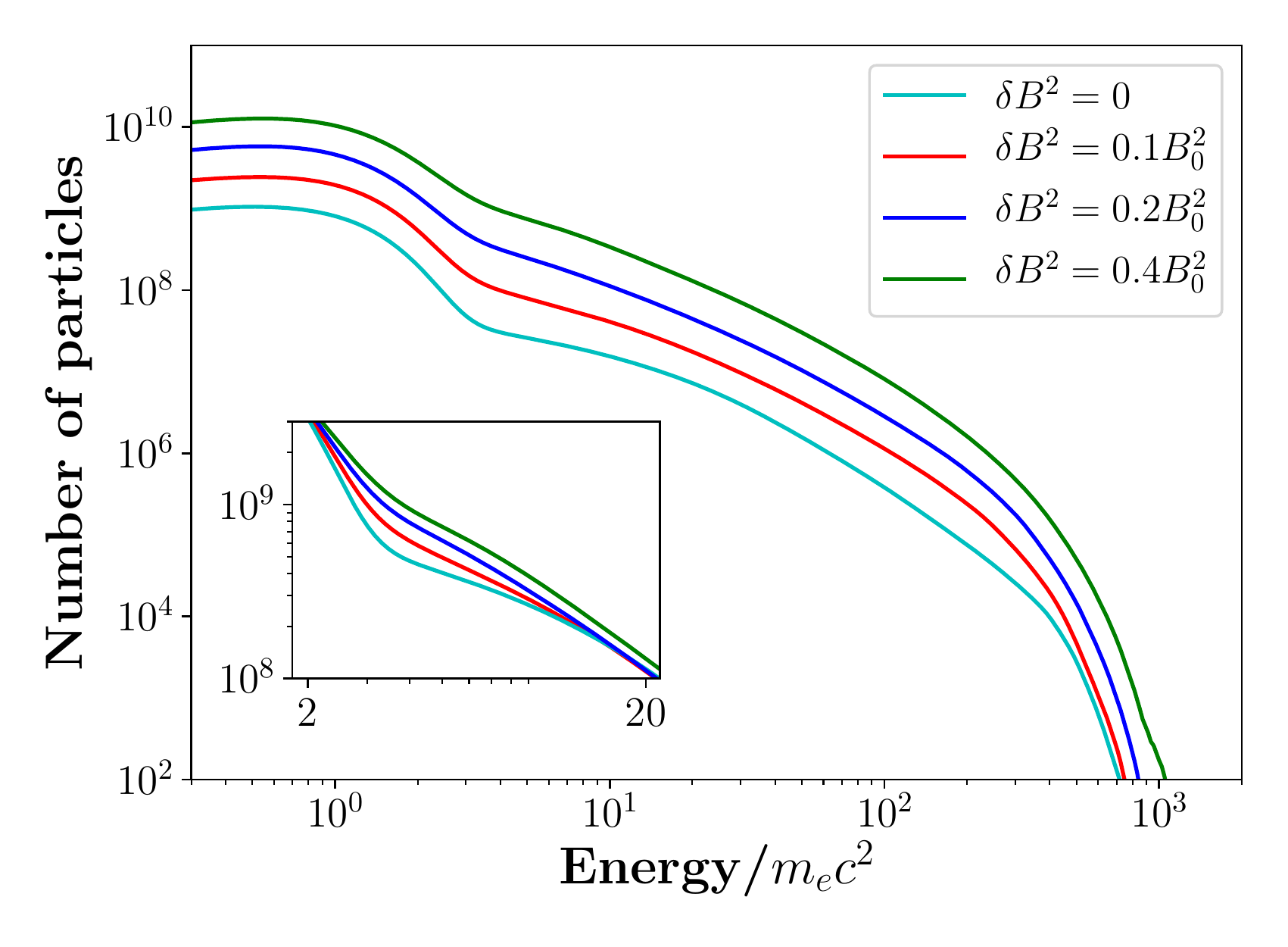}
\caption{\label{fig:spect_deltab}Energy spectra for different turbulence amplitude (slightly shifted up and down against each other). The subpanel compares their absolute flux between $2< \gamma-1<20$. The effect of turbulence accelerates more low energy particles but has not much effect on the high-energy power-law spectrum.}
\end{figure}

Figure~\ref{fig:spect_deltab} shows particle energy spectra for runs with different initial turbulence amplitude. Each of the spectra is slightly shifted up and down to see the difference. One would expect that the injected turbulence contributes to plasma heating and/or particle acceleration. We observe that the flux of the heated part of the distribution $\gamma \lesssim 20$ is increased for cases with higher turbulence injection (shown the subpanel). The thermal core shifts to higher energies as the turbulence amplitude increases because the initial turbulence heats plasma. However, the high-energy spectra above $\gamma > 20$ are nearly identical in terms of flux, spectral index and maximum energy. 
This result shows that the nonthermal particle acceleration is determined by the reconnection dynamics rather than the background turbulence. 


\subsection{Guide field dependence}

\begin{figure}
\centering
\includegraphics[width=0.85\textwidth]{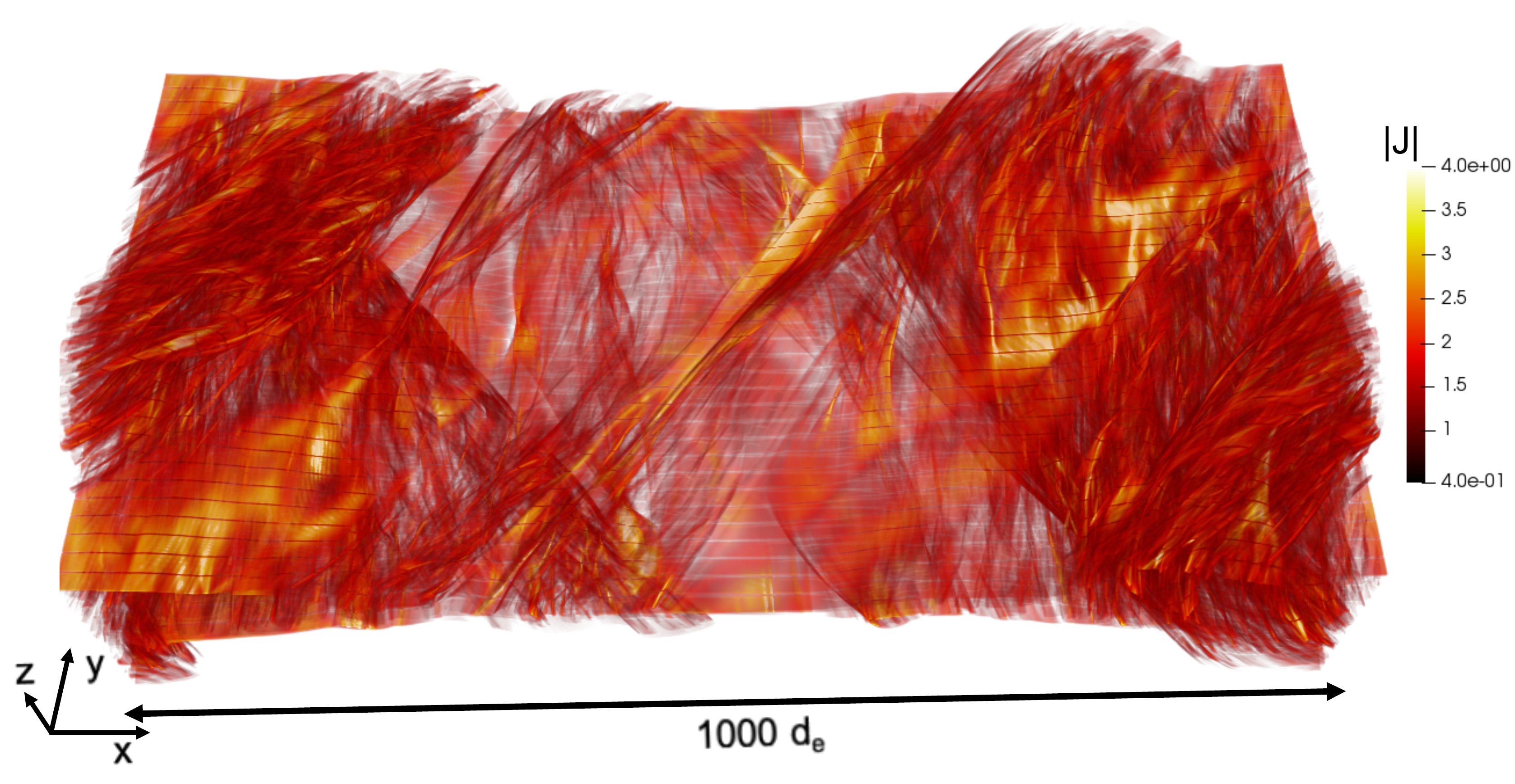}
\caption{\label{fig:absj_bg1}Structure of the reconnection layer for the case with guide field $b_g = 1$ shown using volume rendering of the current density. The reconnection is dominated by flux ropes from the oblique tearing instability.}
\end{figure}

\begin{figure}
\centering
\includegraphics[width=0.6\textwidth]{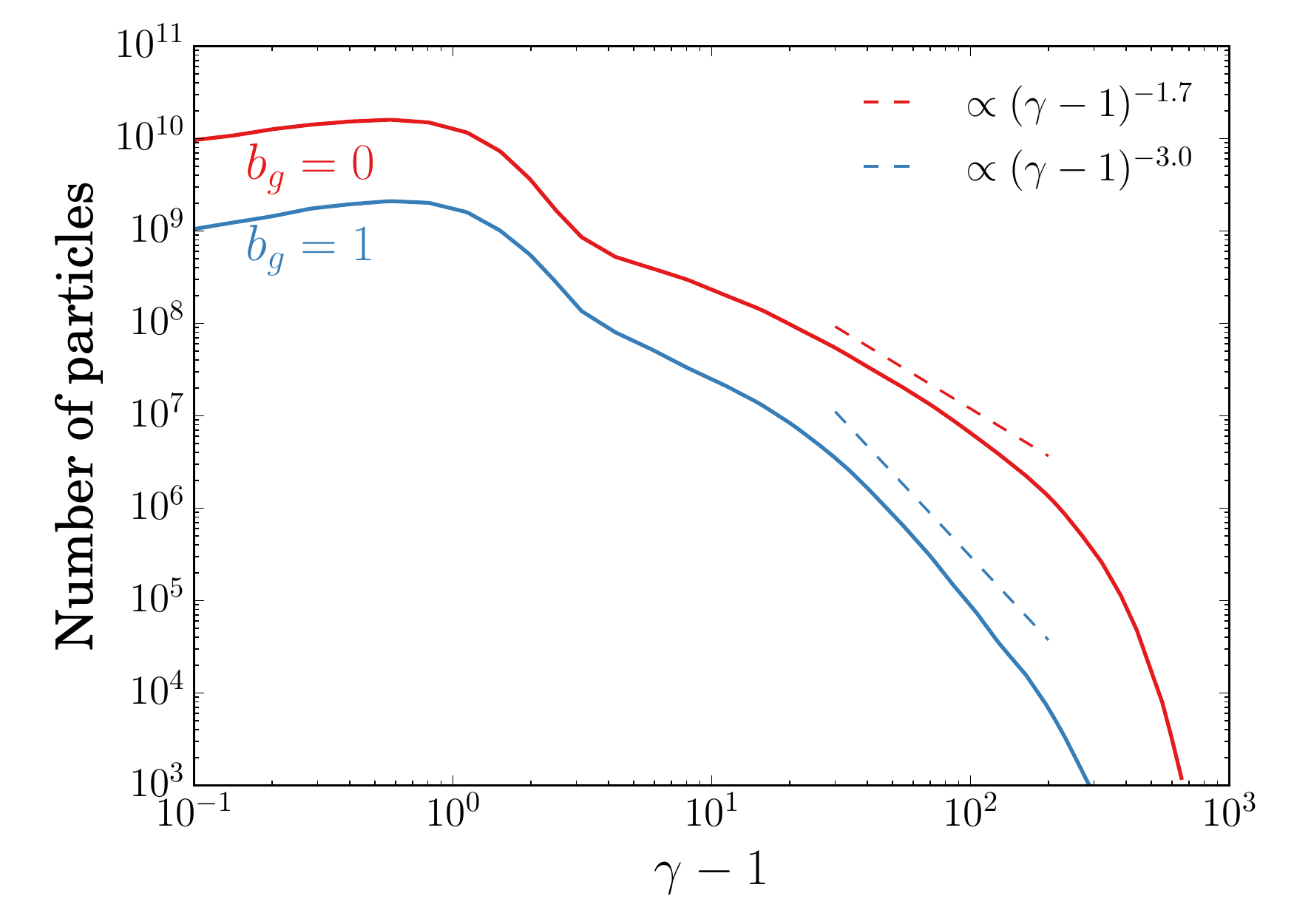}
\caption{\label{fig:spect_bg}Particle energy spectra in the case with $b_g = 1$ (3D-6) compare to the case with $b_g = 0$ (3D-1). In both case $\sigma_e$ calculated based on the reconnecting magnetic field is $100$. The acceleration of particles is less efficient in the presence of a strong guide field.}
\end{figure}

We briefly discuss how a guide field will change reconnection dynamics and particle acceleration processes. Figure~\ref{fig:absj_bg1} shows the structure of the reconnection layer with $b_g=1$, represented by a volume rendering of the current density. One can clearly see the generation of flux ropes oriented obliquely to the guide field direction due to the oblique tearing mode \citep{Daughton2011,Liu2013}. The interaction between flux ropes along different angles leads to a turbulent reconnection layer. The kink mode is suppressed by the guide field \citep{Zenitani2008}. In Figure~\ref{fig:spect_bg}, we show the particle energy spectra in the case with $b_g = 1$ compare to the case with $b_g = 0$. We find that high-energy particle acceleration becomes less efficient and the high-energy spectrum becomes softer, with a spectral index about $p=3.0$.

\begin{figure}
\begin{center}

\includegraphics[width=0.6\textwidth]{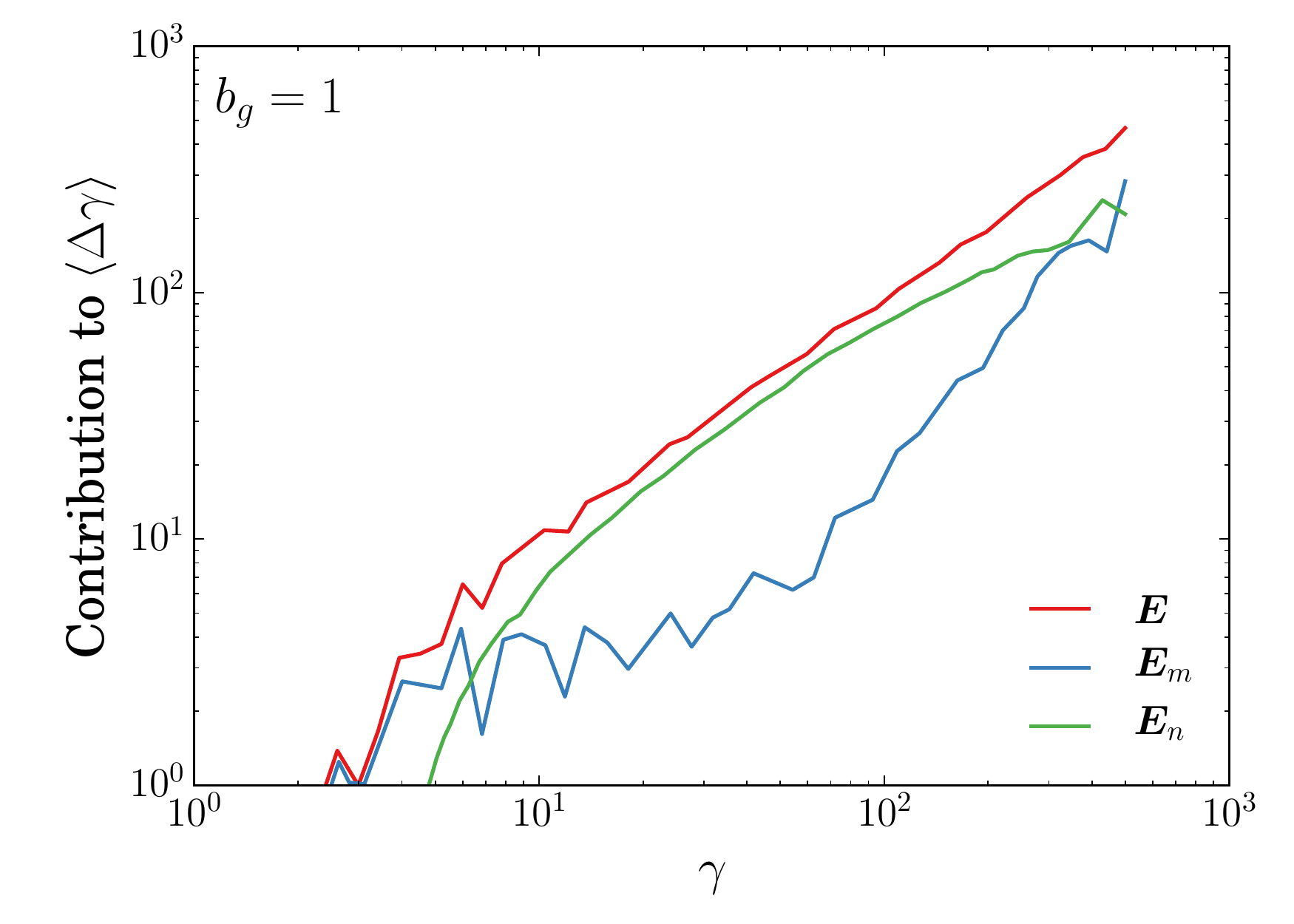}
\caption{\label{fig:egain_bg1}The averaged contribution of the motional electric field $\boldsymbol{E}_m = - \boldsymbol{V} \times \textbf{B} /c$ versus the that of the nonideal electric field $\boldsymbol{E}_n$ to the total energy gain per particle for the case with $b_g = 1$.}
\end{center}
\end{figure}

To understand why the acceleration is less efficient, we perform statistical analysis of acceleration processes and show the results in Figure~\ref{fig:egain_bg1}. We find that the non-ideal acceleration dominates particle acceleration except at the highest energies, which is different from the run when $b_g=0$ (Figure~\ref{fig:egain}). At higher energy the contribution of Fermi acceleration does become comparable to that of the non-ideal electric field acceleration. We would expect that Fermi mechanism will dominate the acceleration processes in larger simulations where particles can be accelerated to higher energies. The relative contributions are similar to the finding in the nonrelativistic case \citep{Dahlin2014,Li2018Role}. \citet{Li2018Role,Li2018Large} have discussed the main controlling physics for this difference \citep[see also][]{leRoux2015}. In the weak guide field and low-$\beta$ (or high $\sigma$) regime, the reconnection layer is highly compressible. This in fact facilitates particle acceleration through a Fermi-like process. When the guide field is stronger, however, the guide field component can prevent the collapse of the reconnection layer, reducing the compressibility and plasmoid formation \citep{Liu2020}, thus the Fermi process is suppressed and the relative contribution of the non-ideal electric field is more important. These results suggest that efficient particle acceleration in magnetic reconnection prefers a weak guide field with $b_g<1$.

\subsection{Superdiffusion of Magnetic Field Lines in the Reconnection Layer}
\label{sec:superdiff}
\begin{figure}
\centering
\includegraphics[width=0.75\textwidth]{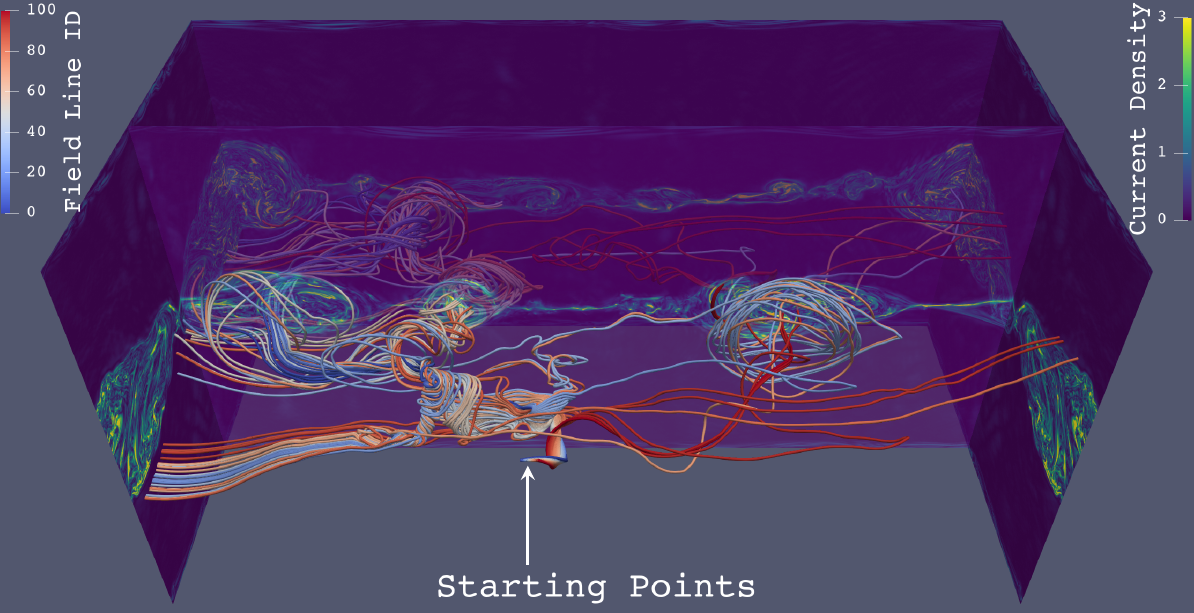}
\caption{Chaotic magnetic field lines starting from 100 points that are uniformly distributed along a line segment with a length of $19d_e$ along $z=0$. The greenish cuts show the current density. The field lines quickly diverge from each other and access the whole simulation domain. Some of the field lines form flux bundles and closely trace the flux ropes. The diverged field lines can also come close to each other again. }
\label{F3}
\end{figure}

In some reconnection models \citep[e.g.,][]{Lazarian1999}, the concept of superdiffusion of magnetic field lines is essential for generating fast turbulent reconnection. Meanwhile, a range of instabilities and kinetic effects have been shown to lead to fast reconnection \citep{Loureiro2007,Daughton2006,Daughton2009,Drake2008,Liu2020}. In our simulations, the reconnection rate stays around $R = 0.1$ while reconnection is in a turbulent state. It is therefore interesting to test some aspects of the superdiffusion concept in our 3D kinetic simulations with physical dimension $L_x / d_e = 1000$.  
Figure~\ref{F3} shows some sample magnetic field lines in the reconnection layer in the standard run. There are 100 field lines started uniformly in a line segment of length of $19 d_e$ along the $x$-direction at $z=0$.
The time snapshot is at 
$\omega_{pe} t = 560$. These field lines are integrated until they reach any simulation boundary.
Because of turbulence in the reconnection layer, field lines can quickly separate from each other and connect to different flux ropes in the simulation.

\begin{figure}
\begin{center}
\includegraphics[width=0.6\textwidth]{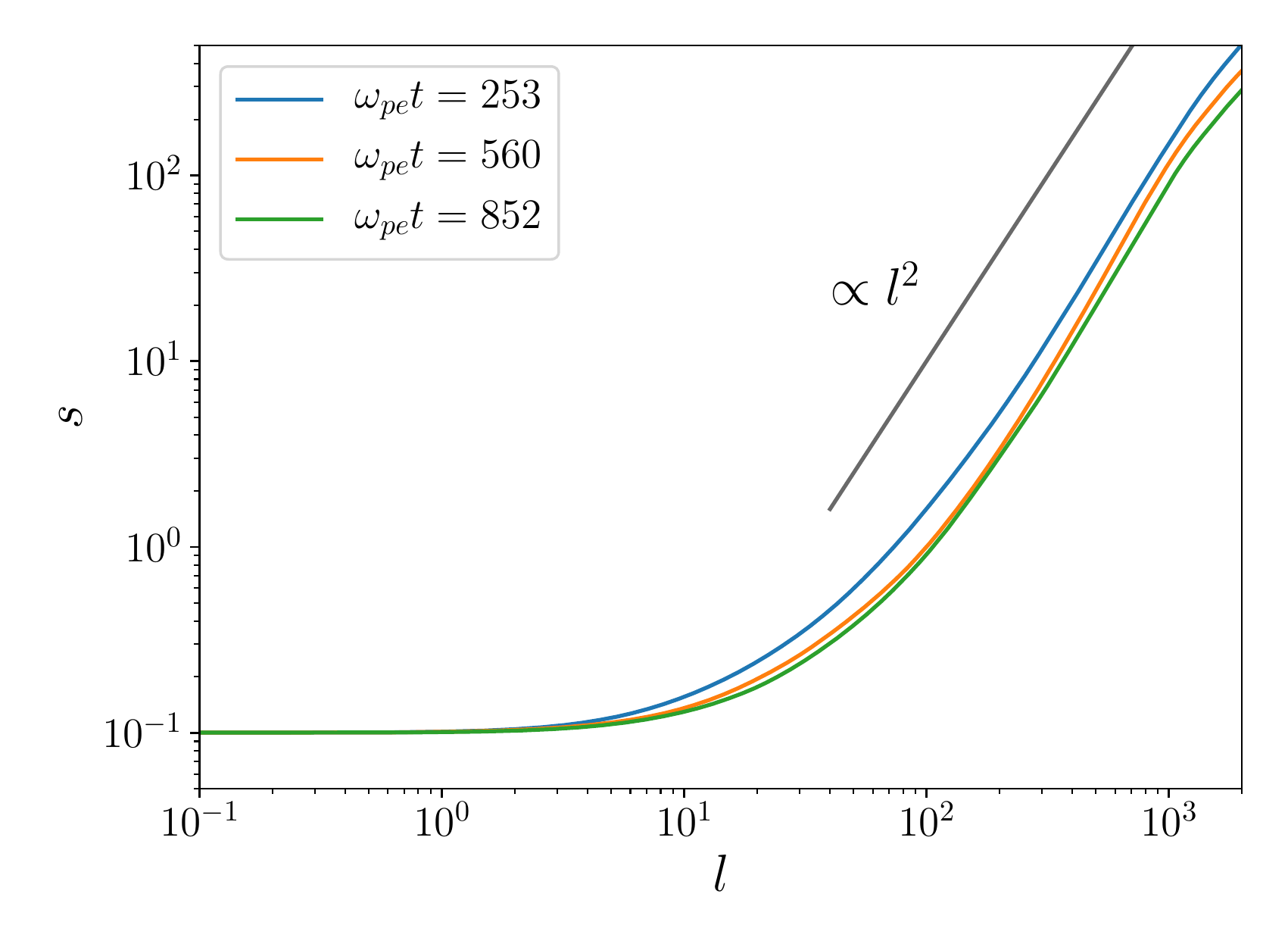}
\end{center}
\caption{Field line diffusion based on the 3D kinetic reconnection simulation run 3D-1.  The three curves in the plots are made using 
three snapshots at time steps corresponding to 
$\omega_{pe} t = 252, 560,$ and $852$, respectively. They represent at-peak, 
post-peak and quasi-steady stages of reconnection, respectively.
They all appear to follow a superdiffusion behavior with $s \propto l^2$.}
\label{F4}
\end{figure}

To quantify the magnetic field diffusion during reconnection, we adopt the following procedure in our simulations: using a magnetic field output at a particular time step, we trace two field lines by starting from a pair of positions that are closely spaced (the typical initial separation is $s_0 = 0.1 d_e$). We calculate the separation between them $s$ as
a function of field line path length $l$. Pairs at different
locations close to the initial current sheet layer center ($z=0$) are chosen
randomly. We have used $10^5$ pairs to enhance the statistics.
Fig. \ref{F4} shows the results of field line separation as a function of fieldline length. At three different time steps $\omega_{pt}t = 253$ (peak reconnection rate)$, 560$ (post-peak) and $852$ (quasi-steady), $s(l)$ follows a relation close to $l^2$ instead of $l^{1/2}$ for diffusion until it starts to roll over at $l \sim 10^3 d_e$. 
Overall, we find that magnetic field lines indeed separate from each other at a rate much
faster than the regular diffusion process. We have chosen two other initial heights 
($z = 2 \lambda$ and $z = 4 \lambda$) for tracing field lines and their behavior is quite similar. 
Furthermore, such analyses were done using three different time-snapshots capturing the
at-peak and post-peak reconnection stages. 
This suggests that magnetic field lines
exhibit super-diffusive behavior. The relatively narrow range of this exponent also suggests that
the initially injected turbulence might have strongly regulated the diffusion behavior,
though the turbulence produced by 3D kinetic reconnection could have impacted the
diffusive dynamics. More detailed analyses are needed to differentiate whether 
the injected and self-generated turbulence could lead to different field line diffusion behavior
and how they might interact with each other. 

While magnetic field lines exhibit super-diffusive 
behavior throughout the reconnection process, the peak reconnection rate does not appreciably depend on the injected turbulence amplitude. The reason for this is not clear (Figure~\ref{fig:rate_deltab}). One possibility is that the reconnection rate in kinetic reconnection is already close to its upper bound due to the force balance in the reconnection region \citep{Liu2017}, and the additional effect of turbulence cannot enhance the rate by any significant factor. This suggests that several different factors, rather than a single mechanism, can contribute to the measured reconnection rate. More effort is needed to identify the main mechanism for fast turbulent reconnection.

\section{Observational Implication}

We now discuss the implication of above results for understanding the role of RTMR in magnetically dominated astrophysical systems. 
It is generally believed that as relativistic jets from black holes or pulsar winds are launched, the flow is dominated by magnetic field with $\sigma \gg 1$. Relativistic magnetic reconnection are likely to present in both relativistic jets and pulsar winds \citep{Giannios2019,Coroniti1990}. The conversion of magnetic energy into particle kinetic energy leads to strong particle energization and high-energy radiation. While earlier 2D studies show that reconnection layer is filled with fast-moving plasmoids that can be approximated as 2D structures, our 3D simulations show a very different picture: RTMR develops in a fully 3D way with three-dimensional instabilities and externally driven turbulence. These 3D features may have a strong impact on high-energy emissions in those systems. Our simulations call into question previous radiation models based on 2D relativistic reconnection \citep{Sironi2016,Petropoulou2016,Zhang2018}, which rely on the more ordered 2D plasmoid structures. Specifically, when 3D effects are considered, the averaged outflow speed becomes much slower; the plasmoid-like structures become quite dynamical and unstable, which cannot be approximated as a cylinder, ellipse, or sphere, as assumed by previous 2D reconnection models. Thus fully 3D radiation modeling is needed to capture the 3D features of RTMR. In the following, we discuss qualitatively the consequences of these 3D features of RTMR on observable signatures.

\subsection{Nonthermal Spectrum}
The overall RTMR nonthermal particle distributions and the resulting radiation spectra are similar to the 2D counterparts.
Generally speaking, the observed high-energy emission from relativistic jets and pulsar winds require the acceleration of a nonthermal power-law energy distribution of particles extending to very high Lorentz factors.
The results presented in this paper further demonstrate that power-law energy spectra of particles are a generic outcome of magnetic reconnection in the magnetically-dominated regime, even when the reconnection process occurs in a turbulent state. The spectral index depends on the magnetization $\sigma_e$ and varies from a soft spectrum ($p=4$ or softer) for small $\sigma_e$ to a hard spectrum when $\sigma_e$ is large. In the limit of large $\sigma_e$ (strong acceleration), the hardest spectrum appears to have $p \sim 1$, which is harder than the usually quoted value for shock acceleration $p \sim 2.2-2.3$ \citep{Achterberg2001,Keshet2005,Yan2016}. In addition, the presence of the guide field can appreciably weaken the acceleration rate, leading to smaller maximum particle energy and softer power-law spectrum.

\subsection{Acceleration Time Scale and Variability}
The current results for relativistic turbulent magnetic reconnection (RTMR) further demonstrate that magnetic reconnection is an efficient mechanism for quickly dissipating magnetic energy in highly magnetized plasmas. The strong radiative cooling and fast flaring activities observed in many high-energy astrophysical systems have suggested the importance of very efficient particle acceleration \citep{Aharonian2007,Abdo2011}, in favor of fast reconnection. Our simulations suggest that external turbulence can be an effective mechanism for triggering magnetic reconnection, leading to a sudden energy release and efficient particle acceleration, consistent with observations. With the existence of turbulence, fast reconnection quickly kicks in and accelerates particles to high energy within a fraction of the light crossing time. However, the strong turbulence and instabilities present in RTMR make the plasmoid-like structures very unstable. As a result, the previous models relying on fast-moving plasmoids to explain fast variability in relativistic jets may be oversimplified.

\subsection{Polarization} 
The strong 3D turbulence and instabilities in RTMR predict very different polarization signatures from the 2D counterparts. Due to the externally applied and/or self-generated turbulence in the reconnection layer, we expect a relatively low polarization degree during reconnection. This can explain the typically observed blazar optical polarization degree at $\sim 10 \%$ level. And {\it IXPE} may also expect relatively low X-ray polarization in the Crab pulsar wind nebula \citep{Weisskopf_2018}. Furthermore, previous 2D reconnection models often simply assume that the plasmoids appear as straight flux ropes or plasma sphere in 3D \citep{Sironi2016,Petropoulou2016,Zhang2018}. As shown in our simulations, the flux ropes are curved and twisted, and can easily get disrupted. Due to the light crossing delay, the light curves, especially at viewing angle other than face-on direction, can appear very different from the 2D simulation results, which demands further studies. Nonetheless, we observe that similar to the 2D plasmoids, the 3D flux ropes can also merge into each other. In addition, these twisted structures can also change in time. Both features can potentially lead to considerable polarization angle rotations at any viewing angles. Obviously, the polarization angle swings are accompanied by blazar flares, due to the strong particle acceleration. Very interestingly, observations have shown that blazar optical angle swings are always simultaneous with {\it Fermi} $\gamma$-ray flares \citep{Blinov2018}. In addition, the polarization degree generally drops during the angle rotations. These features are consistent with the reconnection evolution shown in our simulations, and can be evidence for reconnection in blazar jets.

\section{Conclusion} Thanks to the development of petascale computing and upcoming exascale computers, large-scale particle-in-cell plasma kinetic simulations will allow us to explore 3D plasma dynamics in various processes in a unprecedented way. In this work, we have explored the roles of external turbulent magnetic field on plasma dynamics and particle acceleration in relativistic turbulent magnetic reconnection (RTMR).  We find that during RTMR the current layer breaks up and the reconnection region quickly evolves into a turbulent layer filled with ample coherent structures such as flux ropes and current sheets.  The plasma dynamics in RTMR is quite different from their 2D counterparts in many aspects. The flux ropes evolve rapidly after their generation, and can be completely disrupted due to the secondary kink instability. However, nonthermal particle acceleration and energy release time scale can be very fast and robust, even in the presence of turbulence. We observe clear power-law energy spectra in the magnetically-dominated RTMR regime (from $p \sim 4 $ when $\sigma_e = 6$ to $p \sim 1.3$ when $\sigma_e = 1600$). The main acceleration mechanism for the low-guide-field limit is a Fermi-like acceleration process supported by the motional electric field induced by plasma flows in the reconnection layer, whereas the non-ideal electric field acceleration plays a subdominant role \citep{Litvinenko1996,Sironi2014}. When a significant guide field exists, the kink instability is suppressed and oblique tearing instability becomes the dominant mode that leads to 3D turbulent reconnection \citep{Daughton2011}. In this case the non-ideal electric field can dominate low-energy acceleration, but Fermi acceleration can quickly catch up because its scaling is proportional to energy, suggesting that Fermi acceleration is more important in high-energy acceleration. In addition, we observe that the averaged plasma flow speed in the reconnection layer can be significantly reduced due to the effect of turbulence. These findings have strong implications to high-energy astrophysical systems such as pulsars, jets from black holes, and gamma-ray bursts.

We have also investigated the superdiffusion behavior of magnetic field lines in RTMR. Our analysis suggests that superdiffusion is likely a generic feature of field lines in RTMR. However, for the simulation parameters we explored so far, the reconnection rate is still determined by kinetic physics, as the 3D reconnection rate is similar to its 2D counterpart.

\section{Acknowledgements} 
 We are grateful for support from DOE through the LDRD program at LANL and DoE/OFES support to LANL, and NASA ATP program through grant NNH17AE68I. F. G. and W. D. acknowledge support in part from NASA Grant 80NSSC20K0627. The work by X. L. and Y. L. is funded by the National Science Foundation grant PHY-1902867 through the NSF/DOE Partnership in Basic Plasma Science and Engineering and NASA MMS 80NSSC18K0289. The research by P. K. was also supported by the LANL through its Center for Space and Earth Science (CSES). CSES is funded by LANL's Laboratory Directed Research and Development (LDRD) program under project number 20180475DR. Simulations and analysis were performed on LANL Trinity machine during its open science period. Additional simulations and analysis were performed at National Energy Research Scientific Computing Center (NERSC) and with LANL institutional computing.


\begin{thebibliography}{}
\expandafter\ifx\csname natexlab\endcsname\relax\def\natexlab#1{#1}\fi
\providecommand{\url}[1]{\href{#1}{#1}}
\providecommand{\dodoi}[1]{doi:~\href{http://doi.org/#1}{\nolinkurl{#1}}}
\providecommand{\doeprint}[1]{\href{http://ascl.net/#1}{\nolinkurl{http://ascl.net/#1}}}
\providecommand{\doarXiv}[1]{\href{https://arxiv.org/abs/#1}{\nolinkurl{https://arxiv.org/abs/#1}}}

\bibitem[{{Abdo} {et~al.}(2011){Abdo}, {Ackermann}, {Ajello}, {Allafort},
  {Baldini}, {Ballet}, {Barbiellini}, {Bastieri}, {Bechtol}, {Bellazzini},
  {Berenji}, {Blandford}, {Bloom}, {Bonamente}, {Borgland}, {Bouvier}, {Brand
  t}, {Bregeon}, {Brez}, {Brigida}, {Bruel}, {Buehler}, {Buson}, {Caliandro},
  {Cameron}, {Cannon}, {Caraveo}, {Casand jian}, {{\c{C}}elik}, {Charles},
  {Chekhtman}, {Cheung}, {Chiang}, {Ciprini}, {Claus}, {Cohen-Tanugi},
  {Costamante}, {Cutini}, {D'Ammando}, {Dermer}, {de Angelis}, {de Luca}, {de
  Palma}, {Digel}, {do Couto e Silva}, {Drell}, {Drlica-Wagner}, {Dubois},
  {Dumora}, {Favuzzi}, {Fegan}, {Ferrara}, {Focke}, {Fortin}, {Frailis},
  {Fukazawa}, {Funk}, {Fusco}, {Gargano}, {Gasparrini}, {Gehrels}, {Germani},
  {Giglietto}, {Giordano}, {Giroletti}, {Glanzman}, {Godfrey}, {Grenier},
  {Grondin}, {Grove}, {Guiriec}, {Hadasch}, {Hanabata}, {Harding}, {Hayashi},
  {Hayashida}, {Hays}, {Horan}, {Itoh}, {J{\'o}hannesson}, {Johnson},
  {Johnson}, {Khangulyan}, {Kamae}, {Katagiri}, {Kataoka}, {Kerr},
  {Kn{\"o}dlseder}, {Kuss}, {Lande}, {Latronico}, {Lee}, {Lemoine-Goumard},
  {Longo}, {Loparco}, {Lubrano}, {Madejski}, {Makeev}, {Marelli}, {Mazziotta},
  {McEnery}, {Michelson}, {Mitthumsiri}, {Mizuno}, {Moiseev}, {Monte},
  {Monzani}, {Morselli}, {Moskalenko}, {Murgia}, {Nakamori}, {Naumann-Godo},
  {Nolan}, {Norris}, {Nuss}, {Ohsugi}, {Okumura}, {Omodei}, {Ormes}, {Ozaki},
  {Paneque}, {Parent}, {Pelassa}, {Pepe}, {Pesce-Rollins}, {Pierbattista},
  {Piron}, {Porter}, {Rain{\`o}}, {Rando}, {Ray}, {Razzano}, {Reimer},
  {Reimer}, {Reposeur}, {Ritz}, {Romani}, {Sadrozinski}, {Sanchez},
  {Parkinson}, {Scargle}, {Schalk}, {Sgr{\`o}}, {Siskind}, {Smith}, {Spand re},
  {Spinelli}, {Strickman}, {Suson}, {Takahashi}, {Takahashi}, {Tanaka},
  {Thayer}, {Thompson}, {Tibaldo}, {Torres}, {Tosti}, {Tramacere}, {Troja},
  {Uchiyama}, {Vandenbroucke}, {Vasileiou}, {Vianello}, {Vitale}, {Wang},
  {Wood}, {Yang}, \& {Ziegler}}]{Abdo2011}
{Abdo}, A.~A., {Ackermann}, M., {Ajello}, M., {et~al.} 2011, Science, 331, 739,
  \dodoi{10.1126/science.1199705}

\bibitem[{{Achterberg} {et~al.}(2001){Achterberg}, {Gallant}, {Kirk}, \&
  {Guthmann}}]{Achterberg2001}
{Achterberg}, A., {Gallant}, Y.~A., {Kirk}, J.~G., \& {Guthmann}, A.~W. 2001,
  \mnras, 328, 393, \dodoi{10.1046/j.1365-8711.2001.04851.x}

\bibitem[{{Aharonian} {et~al.}(2007){Aharonian}, {Akhperjanian}, {Bazer-Bachi},
  {Behera}, {Beilicke}, {Benbow}, {Berge}, {Bernl{\"o}hr}, {Boisson}, {Bolz},
  {Borrel}, {Boutelier}, {Braun}, {Brion}, {Brown}, {B{\"u}hler},
  {B{\"u}sching}, {Bulik}, {Carrigan}, {Chadwick}, {Clapson}, {Chounet},
  {Coignet}, {Cornils}, {Costamante}, {Degrange}, {Dickinson},
  {Djannati-Ata{\"\i}}, {Domainko}, {Drury}, {Dubus}, {Dyks}, {Egberts},
  {Emmanoulopoulos}, {Espigat}, {Farnier}, {Feinstein}, {Fiasson},
  {F{\"o}rster}, {Fontaine}, {Funk}, {Funk}, {F{\"u}{\ss}ling}, {Gallant},
  {Giebels}, {Glicenstein}, {Gl{\"u}ck}, {Goret}, {Hadjichristidis}, {Hauser},
  {Hauser}, {Heinzelmann}, {Henri}, {Hermann}, {Hinton}, {Hoffmann}, {Hofmann},
  {Holleran}, {Hoppe}, {Horns}, {Jacholkowska}, {de Jager}, {Kendziorra},
  {Kerschhaggl}, {Kh{\'e}lifi}, {Komin}, {Kosack}, {Lamanna}, {Latham}, {Le
  Gallou}, {Lemi{\`e}re}, {Lemoine-Goumard}, {Lenain}, {Lohse}, {Martin},
  {Martineau-Huynh}, {Marcowith}, {Masterson}, {Maurin}, {McComb}, {Moderski},
  {Moulin}, {de Naurois}, {Nedbal}, {Nolan}, {Olive}, {Orford}, {Osborne},
  {Ostrowski}, {Panter}, {Pedaletti}, {Pelletier}, {Petrucci}, {Pita},
  {P{\"u}hlhofer}, {Punch}, {Ranchon}, {Raubenheimer}, {Raue}, {Rayner},
  {Renaud}, {Ripken}, {Rob}, {Rolland}, {Rosier-Lees}, {Rowell}, {Rudak},
  {Ruppel}, {Sahakian}, {Santangelo}, {Saug{\'e}}, {Schlenker}, {Schlickeiser},
  {Schr{\"o}der}, {Schwanke}, {Schwarzburg}, {Schwemmer}, {Shalchi}, {Sol},
  {Spangler}, {Stawarz}, {Steenkamp}, {Stegmann}, {Superina}, {Tam},
  {Tavernet}, {Terrier}, {van Eldik}, {Vasileiadis}, {Venter}, {Vialle},
  {Vincent}, {Vivier}, {V{\"o}lk}, {Volpe}, {Wagner}, {Ward}, \&
  {Zdziarski}}]{Aharonian2007}
{Aharonian}, F., {Akhperjanian}, A.~G., {Bazer-Bachi}, A.~R., {et~al.} 2007,
  \apjl, 664, L71, \dodoi{10.1086/520635}

\bibitem[{{Arons}(2012)}]{Arons2012}
{Arons}, J. 2012, \ssr, 173, 341, \dodoi{10.1007/s11214-012-9885-1}

\bibitem[{{Beresnyak}(2017)}]{Beresnyak2017}
{Beresnyak}, A. 2017, \apj, 834, 47, \dodoi{10.3847/1538-4357/834/1/47}

\bibitem[{{Biskamp}(2000)}]{Biskamp2000}
{Biskamp}, D. 2000, {Magnetic Reconnection in Plasmas}, Vol.~3 (Springer)

\bibitem[{{Blackman} \& {Field}(1994)}]{Blackman1994}
{Blackman}, E.~G., \& {Field}, G.~B. 1994, \prl, 72, 494,
  \dodoi{10.1103/PhysRevLett.72.494}

\bibitem[{{Blinov} {et~al.}(2018){Blinov}, {Pavlidou}, {Papadakis},
  {Kiehlmann}, {Liodakis}, {Panopoulou}, {Angelakis}, {Balokovi{\'c}},
  {Hovatta}, {King}, {Kus}, {Kylafis}, {Mahabal}, {Maharana}, {Myserlis},
  {Paleologou}, {Papamastorakis}, {Pazderski}, {Pearson}, {Ramaprakash},
  {Readhead}, {Reig}, {Tassis}, \& {Zensus}}]{Blinov2018}
{Blinov}, D., {Pavlidou}, V., {Papadakis}, I., {et~al.} 2018, \mnras, 474,
  1296, \dodoi{10.1093/mnras/stx2786}

\bibitem[{{Bowers} \& {Li}(2007)}]{Bowers2007}
{Bowers}, K., \& {Li}, H. 2007, \prl, 98, 035002,
  \dodoi{10.1103/PhysRevLett.98.035002}

\bibitem[{{Bowers} {et~al.}(2008){Bowers}, {Albright}, {Yin}, {Bergen}, \&
  {Kwan}}]{Bowers2008}
{Bowers}, K.~J., {Albright}, B.~J., {Yin}, L., {Bergen}, B., \& {Kwan},
  T.~J.~T. 2008, Physics of Plasmas, 15, 055703, \dodoi{10.1063/1.2840133}
  
  \bibitem[Cho \& Vishniac(2000)]{Cho2000} Cho, J. \& Vishniac, E.~T.\ 2000, \apj, 539, 273. doi:10.1086/309213


\bibitem[{{Comisso} \& {Asenjo}(2014)}]{Comisso2014}
{Comisso}, L., \& {Asenjo}, F.~A. 2014, \prl, 113, 045001,
  \dodoi{10.1103/PhysRevLett.113.045001}

\bibitem[{{Coroniti}(1990)}]{Coroniti1990}
{Coroniti}, F.~V. 1990, \apj, 349, 538, \dodoi{10.1086/168340}

\bibitem[{{Dahlin} {et~al.}(2014){Dahlin}, {Drake}, \& {Swisdak}}]{Dahlin2014}
{Dahlin}, J.~T., {Drake}, J.~F., \& {Swisdak}, M. 2014, Physics of Plasmas, 21,
  092304, \dodoi{10.1063/1.4894484}

\bibitem[{{Dahlin} {et~al.}(2017){Dahlin}, {Drake}, \& {Swisdak}}]{Dahlin2017}
---. 2017, Physics of Plasmas, 24, 092110, \dodoi{10.1063/1.4986211}

\bibitem[{Daughton \& Karimabadi(2007)}]{Daughton2007}
Daughton, W., \& Karimabadi, H. 2007, Physics of Plasmas, 14, 072303

\bibitem[{{Daughton} {et~al.}(2014){Daughton}, {Nakamura}, {Karimabadi},
  {Roytershteyn}, \& {Loring}}]{Daughton2014}
{Daughton}, W., {Nakamura}, T.~K.~M., {Karimabadi}, H., {Roytershteyn}, V., \&
  {Loring}, B. 2014, Physics of Plasmas, 21, 052307, \dodoi{10.1063/1.4875730}

\bibitem[{{Daughton} {et~al.}(2009){Daughton}, {Roytershteyn}, {Albright},
  {Karimabadi}, {Yin}, \& {Bowers}}]{Daughton2009}
{Daughton}, W., {Roytershteyn}, V., {Albright}, B.~J., {et~al.} 2009, \prl,
  103, 065004, \dodoi{10.1103/PhysRevLett.103.065004}

\bibitem[{{Daughton} {et~al.}(2011){Daughton}, {Roytershteyn}, {Karimabadi},
  {Yin}, {Albright}, {Bergen}, \& {Bowers}}]{Daughton2011}
{Daughton}, W., {Roytershteyn}, V., {Karimabadi}, H., {et~al.} 2011, Nature
  Physics, 7, 539, \dodoi{10.1038/nphys1965}

\bibitem[{Daughton {et~al.}(2006)Daughton, Scudder, \&
  Karimabadi}]{Daughton2006}
Daughton, W., Scudder, J., \& Karimabadi, H. 2006, Physics of Plasmas, 13,
  072101

\bibitem[{{de Gouveia dal Pino} \& {Lazarian}(2005)}]{Pino2005}
{de Gouveia dal Pino}, E.~M., \& {Lazarian}, A. 2005, \aap, 441, 845,
  \dodoi{10.1051/0004-6361:20042590}

\bibitem[{{Drake} {et~al.}(2008){Drake}, {Shay}, \& {Swisdak}}]{Drake2008}
{Drake}, J.~F., {Shay}, M.~A., \& {Swisdak}, M. 2008, Physics of Plasmas, 15,
  042306, \dodoi{10.1063/1.2901194}

\bibitem[{{Drake} {et~al.}(2006){Drake}, {Swisdak}, {Che}, \&
  {Shay}}]{Drake2006}
{Drake}, J.~F., {Swisdak}, M., {Che}, H., \& {Shay}, M.~A. 2006, \nat, 443,
  553, \dodoi{10.1038/nature05116}

\bibitem[{{Giannios} \& {Uzdensky}(2019)}]{Giannios2019}
{Giannios}, D., \& {Uzdensky}, D.~A. 2019, \mnras, 484, 1378,
  \dodoi{10.1093/mnras/stz082}

\bibitem[{{Giannios} {et~al.}(2009){Giannios}, {Uzdensky}, \&
  {Begelman}}]{Giannios2009}
{Giannios}, D., {Uzdensky}, D.~A., \& {Begelman}, M.~C. 2009, \mnras, 395, L29,
  \dodoi{10.1111/j.1745-3933.2009.00635.x}

\bibitem[{{Goldreich} \& {Sridhar}(1995)}]{Goldreich1995}
{Goldreich}, P., \& {Sridhar}, S. 1995, \apj, 438, 763, \dodoi{10.1086/175121}

\bibitem[Goldreich \& Sridhar(1997)]{Goldreich1997} Goldreich, P. \& Sridhar, S.\ 1997, \apj, 485, 680. doi:10.1086/304442


\bibitem[{Guo {et~al.}(2014)Guo, Li, Daughton, \& Liu}]{Guo2014}
Guo, F., Li, H., Daughton, W., \& Liu, Y.-H. 2014, Phys. Rev. Lett., 113,
  155005, \dodoi{10.1103/PhysRevLett.113.155005}

\bibitem[{{Guo} {et~al.}(2019){Guo}, {Li}, {Daughton}, {Kilian}, {Li}, {Liu},
  {Yan}, \& {Ma}}]{Guo2019}
{Guo}, F., {Li}, X., {Daughton}, W., {et~al.} 2019, \apjl, 879, L23,
  \dodoi{10.3847/2041-8213/ab2a15}

\bibitem[{Guo {et~al.}(2015)Guo, Liu, Daughton, \& Li}]{Guo2015}
Guo, F., Liu, Y.-H., Daughton, W., \& Li, H. 2015, The Astrophysical Journal,
  806, 167, \dodoi{10.1088/0004-637x/806/2/167}

\bibitem[Guo et al.(2020)]{Guo2020} Guo, F., Liu, Y.-H., Li, X., et al.\ 2020, Physics of Plasmas, 27, 080501. doi:10.1063/5.0012094


\bibitem[{{Guo} {et~al.}(2016){Guo}, {Li}, {Li}, {Daughton}, {Zhang},
  {Lloyd-Ronning}, {Liu}, {Zhang}, \& {Deng}}]{Guo2016}
{Guo}, F., {Li}, X., {Li}, H., {et~al.} 2016, \apjl, 818, L9,
  \dodoi{10.3847/2041-8205/818/1/L9}

\bibitem[{{Hoshino} \& {Lyubarsky}(2012)}]{Hoshino2012}
{Hoshino}, M., \& {Lyubarsky}, Y. 2012, \ssr, 173, 521,
  \dodoi{10.1007/s11214-012-9931-z}

\bibitem[{{Huang} \& {Bhattacharjee}(2016)}]{Huang2016}
{Huang}, Y.-M., \& {Bhattacharjee}, A. 2016, \apj, 818, 20,
  \dodoi{10.3847/0004-637X/818/1/20}

\bibitem[{{Jokipii}(1973)}]{Jokipii1973}
{Jokipii}, J.~R. 1973, \apj, 183, 1029, \dodoi{10.1086/152289}

\bibitem[{{Karimabadi} {et~al.}(2013){Karimabadi}, {Roytershteyn}, {Daughton},
  \& {Liu}}]{Karamabadi2013}
{Karimabadi}, H., {Roytershteyn}, V., {Daughton}, W., \& {Liu}, Y.-H. 2013,
  \ssr, 178, 307, \dodoi{10.1007/s11214-013-0021-7}

\bibitem[{{Keshet} \& {Waxman}(2005)}]{Keshet2005}
{Keshet}, U., \& {Waxman}, E. 2005, \prl, 94, 111102,
  \dodoi{10.1103/PhysRevLett.94.111102}

\bibitem[Kilian et al.(2020)]{Kilian2020} Kilian, P., Li, X., Guo, F., et al.\ 2020, \apj, 899, 151. doi:10.3847/1538-4357/aba1e9


\bibitem[{{Kirk} \& {Skj{\ae}raasen}(2003)}]{Kirk2003}
{Kirk}, J.~G., \& {Skj{\ae}raasen}, O. 2003, \apj, 591, 366,
  \dodoi{10.1086/375215}
  
  \bibitem[Kowal et al.(2009)]{Kowal2009} Kowal, G., Lazarian, A., Vishniac, E.~T., et al.\ 2009, \apj, 700, 63. doi:10.1088/0004-637X/700/1/63


\bibitem[{{Kowal} {et~al.}(2017){Kowal}, {Falceta-Gon{\c{c}}alves}, {Lazarian},
  \& {Vishniac}}]{Kowal2017}
{Kowal}, G., {Falceta-Gon{\c{c}}alves}, D.~A., {Lazarian}, A., \& {Vishniac},
  E.~T. 2017, \apj, 838, 91, \dodoi{10.3847/1538-4357/aa6001}
  
  \bibitem[Kowal et al.(2020)]{Kowal2020} Kowal, G., Falceta-Gon{\c{c}}alves, D.~A., Lazarian, A., et al.\ 2020, \apj, 892, 50. doi:10.3847/1538-4357/ab7a13


\bibitem[{{Lazarian} \& {Vishniac}(1999)}]{Lazarian1999}
{Lazarian}, A., \& {Vishniac}, E.~T. 1999, \apj, 517, 700,
  \dodoi{10.1086/307233}

\bibitem[{{Leake} {et~al.}(2020){Leake}, {Daldorff}, \& {Klimchuk}}]{Leake2020}
{Leake}, J.~E., {Daldorff}, L. K.~S., \& {Klimchuk}, J.~A. 2020, \apj, 891, 62,
  \dodoi{10.3847/1538-4357/ab7193}
  
  \bibitem[le Roux et al.(2015)]{leRoux2015} le Roux, J.~A., Zank, G.~P., Webb, G.~M., et al.\ 2015, \apj, 801, 112. doi:10.1088/0004-637X/801/2/112


\bibitem[{{Li} {et~al.}(2019{\natexlab{a}}){Li}, {Guo}, \&
  {Li}}]{Li2019Particle}
{Li}, X., {Guo}, F., \& {Li}, H. 2019{\natexlab{a}}, \apj, 879, 5,
  \dodoi{10.3847/1538-4357/ab223b}

\bibitem[{{Li} {et~al.}(2018{\natexlab{a}}){Li}, {Guo}, {Li}, \&
  {Birn}}]{Li2018Role}
{Li}, X., {Guo}, F., {Li}, H., \& {Birn}, J. 2018{\natexlab{a}}, \apj, 855, 80,
  \dodoi{10.3847/1538-4357/aaacd5}

\bibitem[{{Li} {et~al.}(2017){Li}, {Guo}, {Li}, \& {Li}}]{Li2017}
{Li}, X., {Guo}, F., {Li}, H., \& {Li}, G. 2017, \apj, 843, 21,
  \dodoi{10.3847/1538-4357/aa745e}

\bibitem[{{Li} {et~al.}(2018{\natexlab{b}}){Li}, {Guo}, {Li}, \&
  {Li}}]{Li2018Large}
{Li}, X., {Guo}, F., {Li}, H., \& {Li}, S. 2018{\natexlab{b}}, \apj, 866, 4,
  \dodoi{10.3847/1538-4357/aae07b}

\bibitem[{{Li} {et~al.}(2019{\natexlab{b}}){Li}, {Guo}, {Li}, {Stanier}, \&
  {Kilian}}]{Li2019Formation}
{Li}, X., {Guo}, F., {Li}, H., {Stanier}, A., \& {Kilian}, P.
  2019{\natexlab{b}}, \apj, 884, 118, \dodoi{10.3847/1538-4357/ab4268}
  
    \bibitem[Litvinenko(1996)]{Litvinenko1996} Litvinenko, Y.~E.\ 1996, \apj, 462, 997. doi:10.1086/177213

\bibitem[{{Liu} {et~al.}(2013){Liu}, {Daughton}, {Karimabadi}, {Li}, \&
  {Roytershteyn}}]{Liu2013}
{Liu}, Y.-H., {Daughton}, W., {Karimabadi}, H., {Li}, H., \& {Roytershteyn}, V.
  2013, \prl, 110, 265004, \dodoi{10.1103/PhysRevLett.110.265004}

\bibitem[{{Liu} {et~al.}(2015){Liu}, {Guo}, {Daughton}, {Li}, \&
  {Hesse}}]{Liu2015}
{Liu}, Y.-H., {Guo}, F., {Daughton}, W., {Li}, H., \& {Hesse}, M. 2015, \prl,
  114, 095002, \dodoi{10.1103/PhysRevLett.114.095002}

\bibitem[{{Liu} {et~al.}(2017){Liu}, {Hesse}, {Guo}, {Daughton}, {Li},
  {Cassak}, \& {Shay}}]{Liu2017}
{Liu}, Y.-H., {Hesse}, M., {Guo}, F., {et~al.} 2017, \prl, 118, 085101,
  \dodoi{10.1103/PhysRevLett.118.085101}

\bibitem[{{Liu} {et~al.}(2020){Liu}, {Lin}, {Hesse}, {Guo}, {Li}, {Zhang}, \&
  {Peery}}]{Liu2020}
{Liu}, Y.-H., {Lin}, S.-C., {Hesse}, M., {et~al.} 2020, \apjl, 892, L13,
  \dodoi{10.3847/2041-8213/ab7d3f}
  



\bibitem[{{Loureiro} \& {Boldyrev}(2018)}]{Loureiro2018}
{Loureiro}, N.~F., \& {Boldyrev}, S. 2018, \apjl, 866, L14,
  \dodoi{10.3847/2041-8213/aae483}

\bibitem[{{Loureiro} {et~al.}(2007){Loureiro}, {Schekochihin}, \&
  {Cowley}}]{Loureiro2007}
{Loureiro}, N.~F., {Schekochihin}, A.~A., \& {Cowley}, S.~C. 2007, Physics of
  Plasmas, 14, 100703, \dodoi{10.1063/1.2783986}

\bibitem[{{Loureiro} {et~al.}(2009){Loureiro}, {Uzdensky}, {Schekochihin},
  {Cowley}, \& {Yousef}}]{Loureiro2009}
{Loureiro}, N.~F., {Uzdensky}, D.~A., {Schekochihin}, A.~A., {Cowley}, S.~C.,
  \& {Yousef}, T.~A. 2009, \mnras, 399, L146,
  \dodoi{10.1111/j.1745-3933.2009.00742.x}

\bibitem[{{Lyubarsky} \& {Kirk}(2001)}]{Lyubarsky2001}
{Lyubarsky}, Y., \& {Kirk}, J.~G. 2001, \apj, 547, 437, \dodoi{10.1086/318354}

\bibitem[{{Lyubarsky}(2005)}]{Lyubarsky2005}
{Lyubarsky}, Y.~E. 2005, \mnras, 358, 113,
  \dodoi{10.1111/j.1365-2966.2005.08767.x}

\bibitem[{{Lyutikov}(2003)}]{Lyutikov2003}
{Lyutikov}, M. 2003, \mnras, 346, 540, \dodoi{10.1046/j.1365-2966.2003.07110.x}

\bibitem[{{Matthaeus} \& {Lamkin}(1985)}]{Matthaeus1985}
{Matthaeus}, W.~H., \& {Lamkin}, S.~L. 1985, Physics of Fluids, 28, 303,
  \dodoi{10.1063/1.865147}

\bibitem[{{Matthaeus} \& {Lamkin}(1986)}]{Matthaeus1986}
---. 1986, Physics of Fluids, 29, 2513, \dodoi{10.1063/1.866004}

\bibitem[{{McKinney} \& {Uzdensky}(2012)}]{McKinney2012}
{McKinney}, J.~C., \& {Uzdensky}, D.~A. 2012, \mnras, 419, 573,
  \dodoi{10.1111/j.1365-2966.2011.19721.x}
  
  \bibitem[Nathanail et al.(2020)]{Nathanail2020} Nathanail, A., Fromm, C.~M., Porth, O., et al.\ 2020, \mnras, 495, 1549. doi:10.1093/mnras/staa1165


\bibitem[{{Petropoulou} {et~al.}(2016){Petropoulou}, {Giannios}, \&
  {Sironi}}]{Petropoulou2016}
{Petropoulou}, M., {Giannios}, D., \& {Sironi}, L. 2016, \mnras, 462, 3325,
  \dodoi{10.1093/mnras/stw1832}

\bibitem[Priest \& Forbes(2007)]{Priest_2007} Priest, E. \& Forbes, T.\ 2007, Magnetic Reconnection, by Eric Priest , Terry Forbes, Cambridge, UK: Cambridge University Press, 2007


\bibitem[{{Sironi} {et~al.}(2016){Sironi}, {Giannios}, \&
  {Petropoulou}}]{Sironi2016}
{Sironi}, L., {Giannios}, D., \& {Petropoulou}, M. 2016, \mnras, 462, 48,
  \dodoi{10.1093/mnras/stw1620}

\bibitem[{{Sironi} \& {Spitkovsky}(2014)}]{Sironi2014}
{Sironi}, L., \& {Spitkovsky}, A. 2014, \apjl, 783, L21,
  \dodoi{10.1088/2041-8205/783/1/L21}

\bibitem[{Weisskopf(2018)}]{Weisskopf_2018}
Weisskopf, M. 2018, Galaxies, 6, 33, \dodoi{10.3390/galaxies6010033}

\bibitem[{{Werner} \& {Uzdensky}(2017)}]{Werner2017}
{Werner}, G.~R., \& {Uzdensky}, D.~A. 2017, \apjl, 843, L27,
  \dodoi{10.3847/2041-8213/aa7892}

\bibitem[{{Werner} {et~al.}(2016){Werner}, {Uzdensky}, {Cerutti}, {Nalewajko},
  \& {Begelman}}]{Werner2016}
{Werner}, G.~R., {Uzdensky}, D.~A., {Cerutti}, B., {Nalewajko}, K., \&
  {Begelman}, M.~C. 2016, \apjl, 816, L8, \dodoi{10.3847/2041-8205/816/1/L8}

\bibitem[{{Yan} \& {Zhang}(2015)}]{Yan2015}
{Yan}, D., \& {Zhang}, L. 2015, \mnras, 447, 2810,
  \dodoi{10.1093/mnras/stu2551}
  
  \bibitem[Yan et al.(2016)]{Yan2016} Yan, D., He, J., Liao, J., et al.\ 2016, \mnras, 456, 2173. doi:10.1093/mnras/stv2829


\bibitem[Yang et al.(2020)]{Yang2020} Yang, L., Li, H., Guo, F., et al.\ 2020, \apjl, 901, L22. doi:10.3847/2041-8213/abb76b


\bibitem[{{Zenitani} \& {Hoshino}(2008)}]{Zenitani2008}
{Zenitani}, S., \& {Hoshino}, M. 2008, \apj, 677, 530, \dodoi{10.1086/528708}

\bibitem[{{Zhang} \& {Yan}(2011)}]{Zhang2011}
{Zhang}, B., \& {Yan}, H. 2011, \apj, 726, 90,
  \dodoi{10.1088/0004-637X/726/2/90}

\bibitem[{{Zhang} {et~al.}(2015){Zhang}, {Chen}, {B{\"o}ttcher}, {Guo}, \&
  {Li}}]{Zhang2015}
{Zhang}, H., {Chen}, X., {B{\"o}ttcher}, M., {Guo}, F., \& {Li}, H. 2015, \apj,
  804, 58, \dodoi{10.1088/0004-637X/804/1/58}

\bibitem[{{Zhang} {et~al.}(2017){Zhang}, {Li}, {Guo}, \& {Taylor}}]{Zhang2017}
{Zhang}, H., {Li}, H., {Guo}, F., \& {Taylor}, G. 2017, \apj, 835, 125,
  \dodoi{10.3847/1538-4357/835/2/125}

\bibitem[{{Zhang} {et~al.}(2018){Zhang}, {Li}, {Guo}, \&
  {Giannios}}]{Zhang2018}
{Zhang}, H., {Li}, X., {Guo}, F., \& {Giannios}, D. 2018, \apjl, 862, L25,
  \dodoi{10.3847/2041-8213/aad54f}
  
  \bibitem[Zhang et al.(2020)]{Zhang2020} Zhang, H., Li, X., Giannios, D., et al.\ 2020, \apj, 901, 149. doi:10.3847/1538-4357/abb1b0

  
  \bibitem[Zhang \& Giannios(2021)]{Zhang2021a} Zhang, H. \& Giannios, D.\ 2021, \mnras, 502, 1145. doi:10.1093/mnras/stab008
  
  \bibitem[Zhang et al.(2021)]{Zhang2021b} Zhang, Q., Guo, F., Daughton, W., et al.\ 2021, arXiv:2105.04521



\bibitem[{{Zhou} {et~al.}(2020){Zhou}, {Loureiro}, \& {Uzdensky}}]{Zhou2020}
{Zhou}, M., {Loureiro}, N.~F., \& {Uzdensky}, D.~A. 2020, Journal of Plasma
  Physics, 86, 535860401, \dodoi{10.1017/S0022377820000641}

\end{thebibliography}
\bibliographystyle{aasjournal}

\end{document}